\documentclass[12pt]{article}

\usepackage{newtxtext,newtxmath}
\usepackage{graphicx}
\usepackage[letterpaper,margin=1in]{geometry}
\date{}

\makeatletter
\renewcommand{\fnum@figure}{\textbf{Fig. \thefigure}}
\renewcommand{\fnum@table}{\textbf{Table \thetable}}
\makeatother

\usepackage{scicite}
\usepackage{booktabs} 

\usepackage{amsmath,amssymb}

\title{\bfseries \boldmath Temporal magnetic interfaces reveal damping-induced spin-wave amplification near the stripe-domain transition in ultrathin films with DMI}

\author{
Krzysztof Sobucki$^{1}$, Pawel Gruszecki$^{1\ast}$\and
\small$^{1}$Institute of Spintronics and Quantum Information, Faculty of Physics, Adam Mickiewicz University, Pozna\'{n}, Poland\and
\small$^{\ast}$Corresponding author. Email: gruszecki@amu.edu.pl
}

\begin{document}
\maketitle

\begin{abstract}
Using micromagnetic simulations and analytical theory, we study temporal magnetic interfaces in ultrathin CoFeB films with perpendicular magnetic anisotropy and interfacial Dzyaloshinskii--Moriya interaction. We show that time refraction and reflection are governed by precession ellipticity, acting as a magnonic temporal impedance, while smooth field ramps suppress temporal reflections. Near the transition from a uniform state to stripe domains, the exceptional-point and critical fields delimit damping, slow-instability, and strong-instability regimes. In the slow-instability window, Gilbert damping counterintuitively drives spin-wave growth with a rate proportional to the damping parameter. Micromagnetic simulations confirm that a temporal-slab protocol exploiting this regime achieves up to 175-fold frequency-preserving amplitude amplification without continuous power injection. Energy analysis indicates that the field ramp stores energy in the metastable uniform state below the stripe-domain transition, later released into growing spin-wave excitations, consistent with the antimagnonic framework. These results establish temporal field modulation as a route to reconfigurable spin-wave gain.
\end{abstract}

\noindent
\section{Introduction}

    Controlling waves through spatial structuring---mirrors, lenses, waveguides---has driven technologies from telecommunications to medical imaging. A fundamentally different approach, manipulating waves by modulating medium properties in time rather than space, has recently transformed photonics\cite{galiffi2022, moussa2023observation, wang2023metasurface}, yet its implementation in other wave platforms remains comparatively less developed. Magnonics, which processes information using spin waves in magnetic materials, offers nanoscale wavelengths, natural nonreciprocity, and rich nonlinear dynamics\cite{chumak2015magnon, chumak2021roadmap}---properties ideally suited for next-generation signal processing. However, Gilbert damping fundamentally limits propagation distances and prohibits cascaded processing without amplification. Existing gain mechanisms---parametric pumping\cite{melkov1999,Serga2010,Rivard2025} and spin--orbit-torque-driven amplifiers\cite{bauer2015,demidov2017}---suffer from narrow bandwidth, threshold behavior, or the need for continuous power injection, with recent breakthroughs achieving 10--50-fold gains\cite{Breitbach2023,Merbouche2024,Nikolaev2025} while still relying on sustained power or threshold-limited processes. Overcoming dissipation without continuous external drive remains the central challenge for practical magnonic signal processing.
    This raises a key question: can net spin-wave gain be achieved without continuous pumping, using a finite-time control protocol rather than sustained power injection and threshold-limited operation?


    Photonics of time-varying media has demonstrated that temporal modulation enables wave manipulation impossible with static structures: time refraction, reflection, and frequency conversion\cite{pacheco2023,vezzoli2018,bacot2016,fink2001,mostafa2024}, antireflection coatings\cite{akbarzadeh2018}, temporal slabs\cite{moussa2023}, photonic time crystals with momentum band gaps\cite{lustig2018,dikopoltsev2022,wang2023,lyubarov2022,Wang2024_PTC}, and topological phases\cite{dutt2020,wang2020temporal,ni2025,rechtsman2013,galiffi2022}. Most recently, non-Foster temporal metastructures have extended these concepts to wave stopping and genuine amplification of electromagnetic waves\cite{PachecoPena2025}. Yet achieving amplification in photonic time-varying systems invariably requires external energy input---whether through periodic temporal modulation\cite{lyubarov2022}, active non-Foster elements\cite{PachecoPena2025}, or nonlinear optical pumping---to overcome intrinsic dielectric losses that universally oppose gain. Whether analogous temporal control can be realized in magnonic systems---and whether it can address the fundamental challenge of spin-wave damping---remains an open question.

    In magnonics, temporal interface control remains systematically unexplored. Non-Hermitian physics, and in particular exceptional points (EPs), has recently gained attention in magnonic systems~\cite{Yu2024_NonHermitian}.
    EPs are spectral degeneracies at which two or more eigenvalues and their associated eigenvectors simultaneously coalesce, in contrast to ordinary (diabolic) Hermitian degeneracies, where only the eigenvalues coincide~\cite{Heiss2012,Kato1966}.
    Yet isolated demonstrations of temporal phenomena have not been unified into a coherent framework: time reversal~\cite{chumak2010}, experimental time refraction~\cite{Rezende1967,Auld1967,schultheiss2021}, numerical time reflection and refraction~\cite{toedt2021}, Floquet states in magnetic vortices~\cite{heins2026,devolder2025,philippe2025}, and space-time periodic magnetization patterns~\cite{traeger2021}.
    On the gain side, recent amplification breakthroughs via spin-orbit torque~\cite{Merbouche2024} and parametric pumping~\cite{Nikolaev2025,elyasi2022} rely on time-modulated external drives rather than on abrupt changes of the intrinsic medium parameters.
    A distinct route to gain is offered by the recently introduced concept of \emph{antimagnons}---spin-wave modes whose creation lowers, rather than raises, the magnetic energy~\cite{Harms2024antimagnonics,Wang2026antimagnonObs}---which operates in metastable magnetic configurations, but whose connection to temporal-interface scattering remains unexplored.
    A promising avenue emerges from magnetic thin films with perpendicular magnetic anisotropy (PMA) and interfacial Dzyaloshinskii--Moriya interaction (DMI), which undergo field-driven phase transitions between uniform magnetization and modulated magnetic textures such as stripe domains\cite{Kisielewski2023,cepeda2026, contreras2025}. Near this critical point, the spin-wave dispersion softens dramatically---a universal precursor to the emergence of periodic domain patterns\cite{contreras2025, cepeda2026}. The proximity to such phase transitions, combined with DMI-induced nonreciprocity, suggests unexplored dynamical regimes where dissipation and instability may interact in ways fundamentally distinct from photonic platforms.

    Here, using a combination of analytical theory and systematic micromagnetic simulations, we show that Gilbert damping---conventionally the primary obstacle to spin-wave propagation---can enable spin-wave amplification near field-driven phase transitions in ultrathin CoFeB films with perpendicular anisotropy and interfacial DMI. We develop an analytical description based on magnonic temporal impedance and validate it with micromagnetic simulations,
    showing that adiabatic field ramps suppress temporal reflections exponentially, analogous to Landau--Zener dynamics\cite{landau1932theorie, zener1932non}. We identify three dynamical regimes---damping, slow instability, and strong instability---controlled by proximity to an exceptional point and the critical field. Near this exceptional point, finite Gilbert damping lifts the degeneracy and, remarkably, enables exponential amplitude growth: in the slow-instability regime, the growth rate scales linearly with $\alpha$, so increased dissipation strengthens rather than opposes gain. A temporal-slab protocol operating in this regime achieves 175-fold amplitude amplification without continuous power injection. 
    This dissipation-driven gain mechanism is specific to magnetic systems near phase transitions and demonstrates that temporal field modulation provides a practical route to reconfigurable spin-wave amplification.
    Energy analysis shows that the field ramp prepares a metastable uniform state below $H_c$, whose excess energy is subsequently released into the growing spin-wave amplitude, consistent with the recently introduced antimagnonic framework~\cite{Harms2024antimagnonics,Wang2026antimagnonObs}.
    Beyond the specific CoFeB system studied here, the analytical expressions depend only on material parameters, suggesting applicability to other ferromagnetic thin films with PMA and DMI near field-driven phase transitions.

\begin{figure}[htbp]
        \centering
        \includegraphics[width=0.95\textwidth]{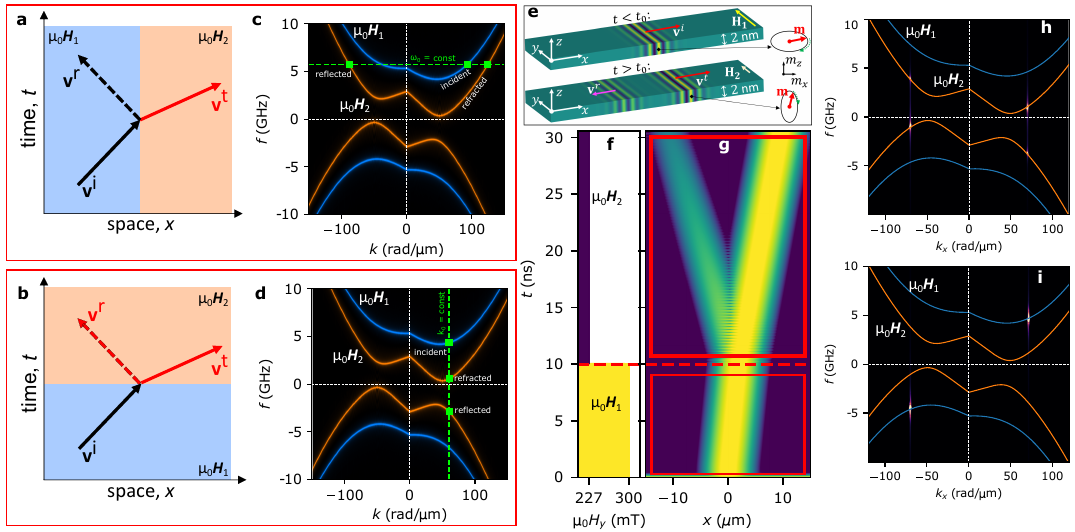}    
        \caption{\textbf{Temporal versus spatial interface scattering in spin-wave systems.}
        \textbf{a,b}, Schematic comparison of wave scattering at spatial (a) and temporal (b) interfaces between media with different dispersion relations, set by different external magnetic fields $H_1$ and $H_2$.
        \textbf{c,d}, Scattering processes in $(f, k)$-space with nonreciprocal dispersion relations for $\mu_0 H_1 = 300$~mT (blue) and $\mu_0 H_2 = 227$~mT (orange); dispersion relations obtained from micromagnetic simulations as 2D FFT of $m_z(t,x)$ (see Methods). At spatial interfaces (c), frequency is conserved while wavevector changes: reflection reverses $k$ sign and thus the phase velocity sign. At temporal interfaces (d), wavevector is conserved while frequency changes: time-reflection reverses the frequency sign and thus the phase velocity.
        \textbf{e}, Schematics of the CoFeB film ($d = 2$~nm; see Methods) before ($t < t_0$) and after ($t > t_0$) the temporal interface, where the bias field along $y$ switches at $t_0$ from $\mu_0 H_1$ to $\mu_0 H_2$, showing the spin-wave wavepackets and their propagation directions $v^{i,r,t}$. Right: magnetization precession ellipses in the $(m_x, m_z)$ plane depicting change of the ellipticity of precession orbits after change of the field value.
        \textbf{f}, Temporal profile of the external magnetic field showing an instantaneous step at $t = 10$~ns from $\mu_0 H_1 = 300$~mT to $\mu_0 H_2 = 227$~mT.
        \textbf{g}, Space-time evolution of spin-wave amplitude $m_z(t,x)$ (normalized to unity at each time step), showing the incident wavepacket ($v^i > 0$) splitting into time-refracted ($v^t > 0$) and time-reflected ($v^r < 0$) packets.
        \textbf{h,i}, Two-dimensional FFT of regions marked in (g): before the interface (h), one spectral peak on the $H_1$ dispersion; after (i), two peaks at matching $k$ but opposite frequencies on the $H_2$ dispersion, confirming wavevector conservation and frequency reversal.
        }
        \label{fig:fig_1}
    \end{figure}
    
\section{Results}

    \subsection{Reflection and refraction of spin waves at sharp temporal interfaces}

        At a \emph{spatial interface}, medium properties change abruptly across space, creating a boundary between regions with different dispersion relations. Frequency is conserved while wavevector changes, causing an incident wave to split into reflected and refracted waves with opposite signs of the wavevector components normal to the interface (Fig.~\ref{fig:fig_1}a, c).

        At a \emph{temporal interface}, the medium undergoes an abrupt change of its dispersion relation throughout the entire system. Magnetization continuity in time requires wavevector conservation ($\mathbf{k} = \mathrm{const}$) while the angular frequency adjusts to satisfy the new dispersion relation. 
        For an incident rightward-propagating wave with positive phase velocity sign $v^{(i)} > 0$ (defined as $v^{(i)} = \Omega^{(i)}/k$) and angular frequency $\Omega^{(i)} > 0$, two solutions emerge with opposite frequency signs: the transmitted, or time-refracted, wave ($v^{(t)} > 0, \Omega^{(t)} > 0$) and the time-reflected wave ($v^{(r)} < 0, \Omega^{(r)} < 0$) (Fig.~\ref{fig:fig_1}b,d). 
        Here, the superscript $(t)$ labels the time-refracted channel, while $(r)$ labels the time-reflected channel. This scattering mechanism reveals fundamentally different wave physics: temporal interfaces enable frequency conversion while preserving wavevector---the inverse of spatial scattering.

        We demonstrate temporal-interface physics in a 2-nm-thick CoFeB film uniformly magnetized in-plane along the $y$-axis, PMA and interfacial DMI (see Methods for material parameters).
        Figs.~\ref{fig:fig_1}c,d show the spin-wave dispersion relations for $\mu_0 H_1 = 300$~mT and $\mu_0 H_2 = 227$~mT. Interfacial DMI breaks mirror symmetry around $k = 0$ and $f = 0$, enabling nonreciprocal propagation  (different wavelengths for opposite-propagating spin waves at the same frequency)---a key ingredient exploited throughout this work. As the field approaches the critical value $H_c \approx 227.2$~mT, only one dispersion branch softens toward zero frequency. Large PMA can produce a dispersion minimum at low fields in the Damon--Eshbach geometry\cite{Banerjee2017,Lesniewski2025}; we plot both positive and negative frequencies to make the temporal-interface scattering more transparent.

        Micromagnetic simulations of a Gaussian spin-wave packet in a $2$~nm CoFeB film at $\mu_0 H_1 = 300$~mT, subjected at $t = 10$~ns to a global step to $\mu_0 H_2 = 227$~mT, confirm this picture: as shown in Fig.~\ref{fig:fig_1}f--g, the packet splits at the temporal interface into a time-refracted ($v^t>0$) and a time-reflected ($v^r<0$) component at fixed $k$, in agreement with the field-dependent dispersion in Fig.~\ref{fig:fig_1}b,d and with the temporal-interface prediction of wavevector conservation and frequency conversion with sign reversal of the reflected wave (Fig.~\ref{fig:fig_1}h--i).

        We obtain analytic scattering coefficients by solving the linearized Landau--Lifshitz equation within the uniform-mode, thickness-averaged dipolar-field approximation appropriate for ultrathin films and assuming small-amplitude dynamics ($m_x, m_z \ll M_s$). The key insight is that continuity of the magnetization at the temporal boundary couples the two magnetization components through their ellipticity ratio. A sudden field change $H_1 \to H_2$ creates a temporal interface where the wavevector $k$ remains invariant while the frequency adjusts to the new dispersion. Two modes emerge in the final medium: a refracted wave (time-refracted, $\Omega_2^{(t)} > 0$) and a time-reflected wave ($\Omega_2^{(r)} < 0$), with the sign reversal of $\Omega$ being crucial for the scattering phenomenology (see Methods).

        Applying the continuity conditions at the temporal interface yields transmission and reflection coefficients that depend on the precession ellipticity---the ratio of out-of-plane to in-plane magnetization components $\varepsilon^z_i = |m_z^{(i)}|/|m_x^{(i)}| = \sqrt{\omega_{z,i}(k)/\omega_{x,i}(k)}$ in temporal region $i$ (see Eq.~\eqref{eq:omega_x} and \eqref{eq:omega_z} for definitions of $\omega_x$ and $\omega_z$). For the $m_x$ component:
        \begin{equation}
            T_x = \frac{1}{2}\left(1 + \frac{\varepsilon^z_1}{\varepsilon^z_2}\right), \quad R_x = \frac{1}{2}\left(1 - \frac{\varepsilon^z_1}{\varepsilon^z_2}\right),
            \label{eq:T_R_x_eps}
        \end{equation}
        and for the $m_z$ component:
        \begin{equation}
            T_z = \frac{1}{2}\left(1 + \frac{\varepsilon^z_2}{\varepsilon^z_1}\right), \quad R_z = \frac{1}{2}\left(1 - \frac{\varepsilon^z_2}{\varepsilon^z_1}\right).
            \label{eq:T_R_z_eps}
        \end{equation}
        These coefficients depend on reciprocal ellipticity ratios: $\varepsilon^z_1/\varepsilon^z_2$ for $m_x$ versus $\varepsilon^z_2/\varepsilon^z_1$ for $m_z$.
        
        The structure of Eqs.~\eqref{eq:T_R_x_eps}--\eqref{eq:T_R_z_eps} reveals a close analogy with classical wave systems, e.g. electromagnetic transmission at dielectric interfaces depends on impedance $Z_\mathrm{EM} = \sqrt{\mu/\varepsilon}$, while transmission line theory uses $Z_\mathrm{TL} = \sqrt{L/C}$. By direct analogy, we define the \emph{magnonic temporal impedance}:
        \begin{equation}
            Z_i \equiv \varepsilon^z_i = \sqrt{\frac{\omega_{z,i}}{\omega_{x,i}}}.
            \label{eq:magnonic_impedance}
        \end{equation}
        In this notation, the coefficients become:
        \begin{equation}
            T_x = \frac{1}{2}\left(1 + \frac{Z_1}{Z_2}\right), \quad R_x = \frac{1}{2}\left(1 - \frac{Z_1}{Z_2}\right),
            \label{eq:T_R_x}
        \end{equation}
        \begin{equation}
            T_z = \frac{1}{2}\left(1 + \frac{Z_2}{Z_1}\right), \quad R_z = \frac{1}{2}\left(1 - \frac{Z_2}{Z_1}\right),
            \label{eq:T_R_z}
        \end{equation}

        a form directly analogous to photonic and electronic interfaces. Note that the $m_x$ and $m_z$ components depend on reciprocal temporal impedance. The amplitude sum rules $T+R=1$, the sign of $R$, and the relation to energy conservation at this active boundary are discussed in Sec.~S1 of the Supplementary Information.

        We note that the term ``spin-wave impedance'' has previously appeared in the magnonic literature to describe the radiation impedance of inductive antennas~\cite{Vanderveken2022,Bruckner2025},
        which quantifies power transfer between microwave circuits and spin-wave systems. By contrast, the magnonic temporal impedance $Z_i$ introduced here characterizes the intrinsic scattering properties of the spin-wave medium at temporal interfaces, in direct analogy to the electromagnetic wave impedance $Z_\mathrm{EM} = \sqrt{\mu/\varepsilon}$ governing spatial dielectric boundaries.

        The reciprocal structure means $T_x > 1$ when $Z_1 > Z_2$, while $T_z > 1$ when $Z_2 > Z_1$; only one component can be amplified at a given interface. The precession ellipse area scales as $T_S = T_x \cdot T_z$, which always exceeds unity when $Z_1 \neq Z_2$. This indicates a universal phenomenon: \emph{at any temporal interface with temporal impedance mismatch, the precession orbit expands}. In PMA systems near phase transitions, $Z_i$ changes dramatically with field, enabling strategic impedance engineering for amplification.

\begin{figure}[htbp]
            \centering
            \includegraphics[width=0.95\textwidth]{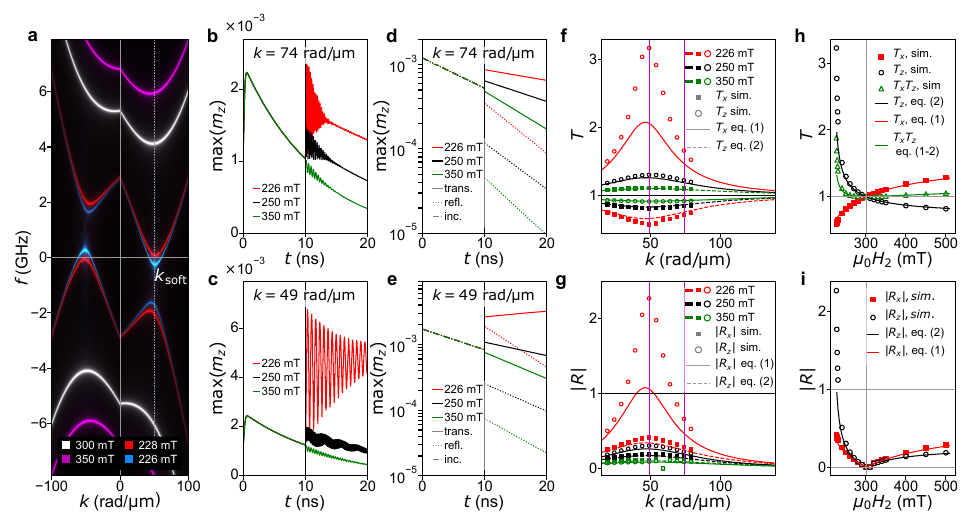}     
            \caption{\textbf{Spin-wave refraction and reflection at sharp temporal magnetic interfaces.}
            \textbf{a}, The spin-wave dispersion relations were extracted from a 2D FFT of the simulated $m_z(t,x)$ data for four different magnetic field values: 350~mT (magenta), 300~mT (white), 228~mT (red), and 226~mT (blue). The grey dashed line marks $k = k_\mathrm{soft}=49$~rad/\textmu m.
            \textbf{b,c}, Time evolution of maximum out-of-plane magnetization component ($\max(m_z)$) for two representative wavenumbers: $k = 74$~rad/\textmu m (b) and $k_\mathrm{soft}$ (c). Spin waves are initially excited at $\mu_0 H = 300$~mT at $t = 0$ and are incident upon a temporal interface at $t = 10$~ns, where the field abruptly changes to $\mu_0 H = 226$~mT (red), $250$~mT (black), or $350$~mT (green). The oscillations of amplitude at $t > 10$~ns arise from the interference between reflected and refracted wave packets.
            \textbf{d,e}, The decomposition of total spin-wave amplitude into incident (dash-dotted), refracted (solid), and reflected (dotted) components, corresponding to panels (b) and (c), respectively, with color coding matching the field values. Decomposition details are in Methods.
            \textbf{f,g}, Wavenumber dependence of the transmission coefficients (f) and reflection coefficients (g) for $\mu_0 H = 226$~mT (red), $250$~mT (black), and $350$~mT (green). Filled squares denote the $x$-component coefficients, $T_x$ in panel f and $|R_x|$ in panel g, whereas open circles denote the $z$-component coefficients, $T_z$ in panel f and $|R_z|$ in panel g. Solid and dashed lines show the corresponding analytical predictions for the $x$ and $z$ components, respectively, from Eqs.~\eqref{eq:T_R_x} and~\eqref{eq:T_R_z}.
            \textbf{h,i}, The magnetic field dependence of transmission (h) and reflection coefficients (i) at fixed $k = k_\mathrm{soft}=49$~rad/\textmu m. Symbols indicate micromagnetic simulation results; lines show analytical model predictions.
            }\label{fig:fig_2}
        \end{figure}
        
        To validate the model, we performed systematic simulations varying $H_2$ and $k$, with initial field $\mu_0 H_1 = 300$~mT.

        Fig.~\ref{fig:fig_2}a summarizes the field-dependent dispersion for $\mu_0 H = 226$--$350$~mT: the dispersion minimum at $k_\mathrm{soft} \approx 49$~rad/\textmu m progressively softens and crosses zero frequency at $\mu_0 H_c \approx 227.2$~mT, signaling the onset of the field-driven transition from the uniform state to the stripe-domain phase. We denote this characteristic wavevector by $k_{\mathrm{soft}}$, defined as the wavevector at which the spin-wave frequency first vanishes at the critical field $H_c$. Although this zero-crossing indicates an incipient instability, the growth of the stripe-domain pattern is sufficiently slow (the origin of this behavior is discussed in the subsequent section) that the uniform state remains metastable over the tens-of-nanoseconds-long window used to extract the dispersion from micromagnetic simulations, even for fields slightly below $H_c$~\cite{Kisielewski2023}. For reference, the dispersion relation in the fully developed, field-aligned stripe-domain configuration at $H < H_c$ is presented in the Supplemental Information (Fig.~S2)~\cite{SM}.

        Figs.~\ref{fig:fig_2}b,c show that a sudden field step generates an exponentially modulated envelope with fast oscillations arising from interference between the refracted and reflected components. The oscillations are strongest and accompanied by net growth at $k_\mathrm{soft}$ and $\mu_0 H_2 = 226$~mT, whereas for other field values the envelope remains nearly constant or decays.

        The decomposition into incident, refracted, and reflected waves' amplitudes (Figs.~\ref{fig:fig_2}d,e) confirms this picture: field steps to lower values ($\mu_0 H_2 = 226, 250$~mT) yield $T_z>1$, while an upward step ($\mu_0 H_2 = 350$~mT) gives $T_z<1$. Reflection is generally subdominant, except for the resonant case ($k_\mathrm{soft}$, $\mu_0 H_2 = 226$~mT), where it becomes comparable and drives strong interference. This identifies $k_\mathrm{soft}$ and $H_2$ just below $H_c$ as the working point where a single temporal step most strongly redistributes amplitude of the incident wave between the refracted and reflected waves, consistent with the peaks of the transmission and reflection coefficients in Figs.~\ref{fig:fig_2}f--i.
        Crucially, beyond the temporal interface the spin-wave amplitude of the refracted wave in Fig.~\ref{fig:fig_2}e continues to grow as the wave propagates in the low-field medium at  $\mu_0 H_2 = 226$~mT---a striking observation that cannot be explained by the temporal scattering coefficients alone. The origin of this sustained amplification is analyzed in the next section, where we show that it arises from a damping-induced instability near the phase transition and constitutes the key mechanism enabling significant gain.

        Figs.~\ref{fig:fig_2}f--i show transmission and reflection coefficients versus $H_2$ and $k$. 
        The transmission coefficients follow a simple monotonic trend with field at fixed $k$: when the field is lowered $T_z$ increases while $T_x$ decreases.
        At $k_\mathrm{soft}$, $T_z$ grows from $T_z = 1$ at $\mu_0 H_2 = 300$~mT to approximately $3$ at $\mu_0 H_2 = 226$~mT, whereas $T_x$ is correspondingly suppressed below unity. The product $T_x T_z > 1$ for all $H_2 \neq 300$~mT confirms the universal expansion of the precession orbit at any temporal impedance mismatch. The reflection coefficients vanish at $\mu_0 H_2 = 300$~mT, and their relative weight swaps across this impedance-matching point: for $H_2 < 300$~mT one finds $|R_z| > |R_x|$, while for $H_2 > 300$~mT the inequality reverses.
        Near $H_c$, $|R_z|$ exhibits a pronounced maximum exceeding $2$ at $\mu_0 H_2 = 226$~mT and $k = k_{\mathrm{soft}}$.

        A good agreement between the analytical predictions and simulations is evident for $H_2 > H_c$, validating that precession ellipticity governs transmission. 
        For $H_2 \lesssim H_c$ (particularly for $\mu_0 H_2 = 226$~mT), the linear model qualitatively captures the resonant features but underestimates the $T_z$ and $|R_z|$ amplitudes and predicts a slightly lower critical field. In micromagnetic simulations the mode softening is more pronounced, yielding a critical field higher by $\sim 4.5$~mT. We attribute this systematic offset and amplitude mismatch to the approximations in the analytical model (uniform-mode, thickness-averaged dipolar field).

    \begin{figure}[htbp]
        \centering
        \includegraphics[width=0.99\textwidth]{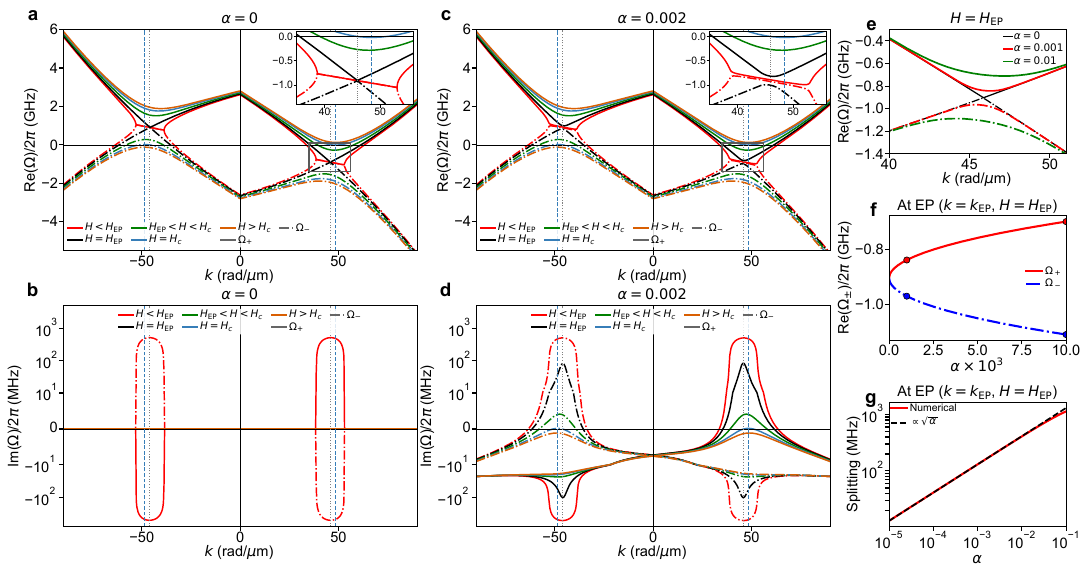}
        \caption{\textbf{Complex spin-wave dispersion and damping-induced splitting near the exceptional point.}
        \textbf{a}, Real part of frequency Re($\Omega$) versus wavevector $k$ for $\alpha = 0$ at five fields: $H < H_{\mathrm{EP}}$ (red), $H = H_{\mathrm{EP}}$ (black), $H_{\mathrm{EP}} < H < H_c$ (green), $H = H_c$ (blue), and $H > H_c$ (orange). Solid and dash-dotted curves denote $\Omega_+$ and $\Omega_-$, respectively. The inset magnifies the exceptional-point region; the rectangle marks the zoomed range. Grey dotted and blue dashed vertical lines indicate $k_{\mathrm{EP}}$ and $k_{\mathrm{soft}}$.
        \textbf{b}, Imaginary part $\mathrm{Im}(\Omega)$ versus $k$ for $\alpha = 0$ on a symmetric logarithmic scale. For $H \geq H_{\mathrm{EP}}$, $\mathrm{Im}(\Omega)=0$, whereas instability occurs for $H < H_{\mathrm{EP}}$.
        \textbf{c}, Same as (a) for $\alpha = 0.002$.
        \textbf{d}, Same as (b) for $\alpha = 0.002$. Finite damping produces non-zero $\mathrm{Im}(\Omega)$ in all regimes; in particular, the green curves show $\mathrm{Im}(\Omega)>0$ in the otherwise stable interval $H_{\mathrm{EP}} < H < H_c$.
        \textbf{e}, Zoom of Re($\Omega$) near $k = k_\mathrm{EP}$ at $H = H_{\mathrm{EP}}$, showing branch splitting for $\alpha = 0.001$ (red) and $\alpha = 0.01$ (green), compared with $\alpha = 0$ (black, degenerate branches).
        \textbf{f}, Re($\Omega_\pm$) at $k = k_{\mathrm{EP}}$ and $H = H_{\mathrm{EP}}$ versus $\alpha$.
        \textbf{g}, Frequency splitting at the exceptional point versus $\alpha$. The numerical result (red) follows the $\sqrt{\alpha}$ scaling (black dashed).
        The exceptional point is located at $\mu_0 H_{\mathrm{EP}} = 219.3$ mT and $k_{\mathrm{EP}} = 45.9$ rad/$\mu$m, with the counterpart at $-k_{\mathrm{EP}}$. For $H = H_{\mathrm{EP}} - 1$ mT, the positive-wavevector exceptional points are $k_{\mathrm{EP},1}=38.7$ and $k_{\mathrm{EP},2}=53.0$ rad/$\mu$m, with corresponding negative-$k$ points. The soft mode occurs at $\mu_0 H_c = 222.5$ mT and $k_{\mathrm{soft}} = 48.6$ rad/$\mu$m.
        }\label{fig:fig_3}
    \end{figure}

    \subsection{Origin of amplification below the critical field}

        The amplification mechanism observed in Fig.~\ref{fig:fig_2}e for waves refracted into the $\mu_0 H_2 = 226$~mT medium follows from the complex frequency $\Omega = \Omega_r + i\Omega_i$: amplitude evolves as $|m| \propto e^{\Omega_i t}$, with positive $\Omega_i$ producing growth.
        The two branches $\Omega_\pm$ are complex eigenvalues of the linearized Landau--Lifshitz--Gilbert equations for the coupled in-plane ($m_x$) and out-of-plane ($m_z$) magnetization components [Eqs.~\eqref{eq:LLG_damped_mx}--\eqref{eq:LLG_damped_mz} of Methods, with full derivation in Section~S3 of the Supplementary Information].
        The real part $\mathrm{Re}(\Omega_\pm)$ is the eigenfrequency of the spin-wave mode, and the imaginary part $\mathrm{Im}(\Omega_\pm)$ sets the growth rate ($\mathrm{Im}>0$) or decay rate ($\mathrm{Im}<0$) of its amplitude.
        Of the parameters entering $\Omega_\pm(\alpha, k, H)$, the Gilbert damping $\alpha$ is fixed by the choice of material, the wavevector $k$ is set by the excitation antenna together with the dispersion relation (which selects the $k$ corresponding to a given excitation frequency), and the external field $H$ is the control parameter swept in time during the temporal protocol.
        Conventionally, Gilbert damping yields $\Omega_i < 0$, requiring external mechanisms for amplification.
        Here, the interplay of damping and DMI creates an intrinsic amplification channel.

\begin{figure}[htbp]
            \centering
            \includegraphics[width=\textwidth]{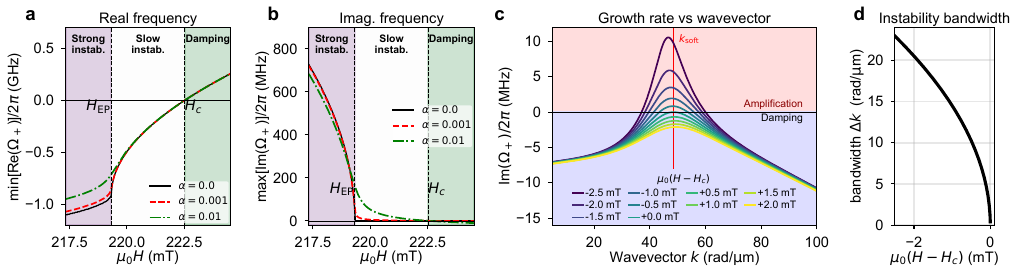}
            \caption{\textbf{Dynamical regimes and wavevector selectivity of temporal amplification.}
            \textbf{a}, Minimum of $\mathrm{Re}(\Omega_+)$ over all wavevectors as a function of magnetic field for three damping values: $\alpha = 0$ (black solid), $\alpha = 0.001$ (red dashed), and $\alpha = 0.01$ (green dash-dotted).
            Background colors indicate the dynamical regimes: strong instability (purple, $H < H_{\mathrm{EP}}$), slow instability (white, $H_{\mathrm{EP}} < H < H_c$), and damping (green, $H > H_c$).
            Vertical dashed lines mark $H_{\mathrm{EP}}$ and $H_c$.
            The condition $\min[\mathrm{Re}(\Omega_+)] = 0$ defines the critical field $H_c$, corresponding to the soft-mode at $k = k_{\mathrm{soft}}$.
            \textbf{b}, Maximum of $\mathrm{Im}(\Omega_+)$ over all wavevectors versus field, with the same color coding as in (a).
            For $\alpha = 0$, the growth rate vanishes throughout the slow-instability regime; finite damping activates a non-zero growth rate in this regime, and larger $\alpha$ yields stronger amplification, demonstrating the counterintuitive enhancement of growth by dissipation.
            \textbf{c}, Growth rate $\mathrm{Im}(\Omega_+)/2\pi$ versus wavevector $k$ for $\alpha = 0.002$ at several values of $\mu_0(H - H_c)$ (labels in mT). Shading indicates amplification (red, $\mathrm{Im}(\Omega_+) > 0$) and damping (blue, $\mathrm{Im}(\Omega_+) < 0$) regions. The growth rate is peaked around $k \approx k_{\mathrm{soft}}$, demonstrating the wavevector selectivity of the temporal amplification mechanism.
            \textbf{d}, Instability bandwidth $\Delta k$, defined as the range of wavevectors for which $\mathrm{Im}(\Omega_+) > 0$, versus $\mu_0(H - H_c)$ for $\alpha = 0.002$.
            The bandwidth vanishes at $H = H_c$ and grows monotonically as the field decreases below $H_c$, reflecting the progressive widening of the unstable wavevector window.
            }
            \label{fig:k_selectivity}
        \end{figure}

        The system dynamics are governed by two characteristic fields (Fig.~\ref{fig:k_selectivity}a,b). 
        We define the critical field $H_c$ as the largest field at which $\min_k[\mathrm{Re}(\Omega_+(k))] = 0$, equivalently the field at which the soft mode at $k = k_{\mathrm{soft}}$ first reaches zero frequency, marking the onset of the field-driven phase transition.
        The lower characteristic field $H_{\mathrm{EP}}$ is defined as the largest field at which $\omega_x(k,H) = 0$ admits a real positive-wavevector solution; equivalently, $H_{\mathrm{EP}} = \max_k h_{\mathrm{EP}}(k)$.
        At $H = H_{\mathrm{EP}}$, these two solutions merge into a single exceptional point at $k_{\mathrm{EP}} = 45.9$ rad/$\mu$m, with the symmetry-related counterpart at $-k_{\mathrm{EP}}$. For $H < H_{\mathrm{EP}}$, this double root unfolds into two positive-wavevector exceptional points; for example, at $H = H_{\mathrm{EP}} - 1$ mT we find $k_{\mathrm{EP},1}=38.7$ and $k_{\mathrm{EP},2}=53.0$ rad/\textmu m, together with their negative-$k$ counterparts. 
        In the conservative limit ($\alpha = 0$), the exceptional point corresponds to the coalescence of both eigenvalues and eigenvectors. For finite damping, this degeneracy is lifted with a splitting $|\Omega_+ - \Omega_-| \propto \sqrt{\alpha}$ (Fig.~\ref{fig:fig_3}e--g). Since $\omega_x^*>0$, one always has $H_{\mathrm{EP}} < H_c$.


        These two fields delineate three distinct dynamical regimes presented in Table~\ref{tab:regimes_results} and Fig.~\ref{fig:k_selectivity}a,b.

    \begin{table}[!b]
            \centering
            \caption{\textbf{Dynamical regimes of spin-wave amplification.}
                $H_c$ denotes the critical field (phase transition onset, $\omega_x = \omega_x^*$); $H_{\mathrm{EP}}$ marks the exceptional point ($\omega_x = 0$).
                In the damping and slow-instability regimes the rate scales linearly with $\alpha$, whereas in the strong-instability regime it is set primarily by $\sqrt{|\omega_x|\omega_z}$ (cf. Eq.~\eqref{eq:Im_strong_instability}).}
            \label{tab:regimes_results}
            \begin{tabular}{@{}lcccc@{}}
                \toprule
                Regime & Field range & $\mathrm{Re}(\Omega)$ & $\mathrm{Im}(\Omega)$ & Amplitude dynamics \\
                \midrule
                Damping & $H > H_c$ & $> 0$ & $< 0$ & Decay, rate $\propto \alpha$ \\[4pt]
                Slow instability & $H_{\mathrm{EP}} < H < H_c$ & $< 0$ & $> 0$ & Growth, rate $\propto \alpha$
                \\[4pt]
                Strong instability & $H < H_{\mathrm{EP}}$ & $< 0$ & $> 0$ & Growth, rate $\sim \sqrt{|\omega_x|\omega_z}$ \\
                \bottomrule
            \end{tabular}
        \end{table}

        The comparison between $\alpha = 0$ and $\alpha > 0$ cases reveals the central result (Fig.~\ref{fig:fig_3}a--d). For $\alpha = 0$, the frequency is purely real when $\omega_x > 0$ (Fig.~\ref{fig:fig_3}a); instability ($\mathrm{Im}(\Omega) > 0$) occurs only for $H < H_{\mathrm{EP}}$ where the uniform magnetic configuration itself is unstable (Fig.~\ref{fig:fig_3}b, red curves). 
        In stark contrast, for $\alpha > 0$, amplification emerges in the range $H_{\mathrm{EP}} < H < H_c$---a regime that would be completely stable in the absence of damping (Fig.~\ref{fig:fig_3}d, green curves showing $\mathrm{Im}(\Omega) > 0$). 

        The physical origin lies in the proximity to the exceptional point. At $\omega_x = 0$ in the limit $\alpha = 0$, both dispersion branches coalesce: $\Omega_+ = \Omega_- = -s\omega_D(k)$ (Fig.~\ref{fig:fig_3}a, black curves at $H = H_{\mathrm{EP}}$), and the dynamical matrix becomes non-diagonalizable (see proof in SM). Crucially, when damping is introduced as shown in Fig.~\ref{fig:fig_3}e-g, this degeneracy is lifted with splitting $|\Omega_+ - \Omega_-| \propto \sqrt{\alpha}$---the hallmark of EP physics, distinct from linear splitting at ordinary degeneracies. 
        This lifting of degeneracy is precisely what creates the slow instability regime: the two branches that touch at $H = H_{\mathrm{EP}}$ for $\alpha = 0$ acquire distinct imaginary parts for $\alpha > 0$, with one branch ($\Omega_+$) gaining a positive imaginary component in the range $H_{\mathrm{EP}} < H < H_c$. Thus, the damping-induced splitting of the exceptional point directly enables amplification in a field window that would be completely stable without dissipation. 
        Exceptional points and dissipation-induced instabilities have been extensively explored in photonic systems\cite{Miri2019_EP}, yet the specific realization demonstrated here---coupling Gilbert damping to DMI-mediated nonreciprocity near a field-driven magnetic phase transition---represents a distinctly magnonic pathway to amplification.
        This represents an amplification mechanism where dissipation enables gain, a counterintuitive behavior that, while conceptually related to loss-induced phenomena in non-Hermitian photonics, emerges here from the specific interplay of Gilbert damping and magnetic phase transition dynamics.
        In practical terms, a small field window opens up between the onset of the strong instability and the phase transition, where spin waves that would normally decay instead grow exponentially because damping lifts the exceptional-point degeneracy.

        In the damping regime ($H > H_c$, green background in Fig.~\ref{fig:k_selectivity}a,b), the decay rate scales linearly with $\alpha$. In the slow instability regime ($H_{\mathrm{EP}} < H < H_c$, white background), the growth rate also scales linearly with $\alpha$---this damping-induced amplification vanishes in the conservative limit $\alpha \to 0$. Remarkably, these two regimes are governed by a single unified expression (see Methods, Eq.~\eqref{eq:Im_unified}): $\mathrm{Im}(\Omega_+) \propto \alpha[\sqrt{\omega_x^*/\omega_x} - 1]$, where the sign is determined solely by whether $\omega_x$ lies below ($\mathrm{Im} > 0$, amplification) or above ($\mathrm{Im} < 0$, damping) the critical value $\omega_x^* = \omega_D^2/\omega_z$. In the strong instability regime ($H < H_{\mathrm{EP}}$, purple background), growth is governed by the magnetic instability with a dominant $\alpha$-independent term $\sqrt{|\omega_x|\omega_z}$.

        Counterintuitively, larger damping enhances amplification in the slow instability regime since the growth rate $\mathrm{Im}(\Omega)$ is proportional to $\alpha$ when $\omega_x < \omega_x^*$. This is evident in Fig.~\ref{fig:k_selectivity}b, where larger $\alpha$ (green curve, $\alpha = 0.01$) produces higher $\mathrm{Im}(\Omega)$ values compared to smaller $\alpha$ (red curve, $\alpha = 0.001$), and is confirmed by micromagnetic simulations discussed later in this work showing $T \propto \exp(\mathrm{const}\cdot \alpha)$ (Fig.~\ref{fig:fig_6}e,f). The amplification window $\Delta H = H_c - H_{\mathrm{EP}}$ spans approximately $3$--$4$~mT for our CoFeB parameters (Fig.~\ref{fig:k_selectivity}a). Importantly, this width scales as $D^2/M_s^3$ and is independent of $\alpha$ (see Methods), while damping controls only the growth rate within the window. Consequently, standard ferromagnetic materials with moderate damping ($\alpha \sim 0.01$) are well-suited for this mechanism---the damping enhances gain without narrowing the operational field range.


        The same expression for $\mathrm{Im}(\Omega_+)$ [Eq.~\eqref{eq:Im_unified}] also predicts a sharp wavevector selectivity of the amplification mechanism. Fig.~\ref{fig:k_selectivity}c shows the growth rate $\mathrm{Im}(\Omega_+)/2\pi$ versus wavevector for several fields near $H_c$.
        Close to the critical field, the growth rate is sharply peaked near $k_{\mathrm{soft}}$, while the uniform mode ($k \to 0$) remains stable. Lowering the field below $H_c$ both enhances the peak gain and broadens the instability bandwidth $\Delta k$, defined as the range of wavevectors with $\mathrm{Im}(\Omega_+) > 0$ (Fig.~\ref{fig:k_selectivity}d). This intrinsic $k$-selectivity provides natural filtering: only modes near $k_{\mathrm{soft}}$ experience significant amplification, while other wavevectors are damped. The selectivity can be tuned by material parameters---shifting $k_{\mathrm{soft}}$ through PMA or DMI engineering---enabling application-specific bandwidth design.
        
    \begin{figure}[htbp]
        \centering
        \includegraphics[width=0.95\textwidth]{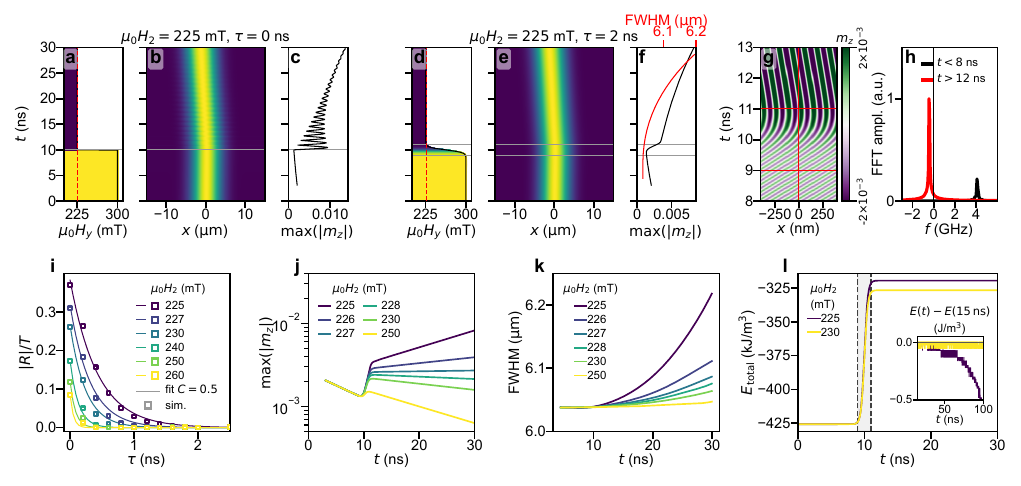}
        \caption{\textbf{Temporal interfaces and spin-wave response to field modulation.}
        \textbf{a--f}, Comparison of spin-wave dynamics for abrupt and finite-duration temporal interfaces. In each set of three panels, the left panel shows the time-dependent magnetic field $\mu_0 H_y(t)$, the middle panel shows the space--time evolution of the spin-wave amplitude $m_z(t,x)$, normalized at each time step, and the right panel shows $\max(|m_z|)$ versus time. In panel \textbf{f}, the red curve on the upper axis additionally shows the FWHM of the wavepacket. Panels \textbf{a--c} correspond to an abrupt field change ($\tau=0$) from $\mu_0 H_1=300$~mT to $\mu_0 H_2=225$~mT, whereas panels \textbf{d--f} correspond to a finite temporal interface ($\tau=2$~ns) for the same field values.
        \textbf{g}, Zoomed space--time magnetization map from (e) around the temporal interface, showing the change in frequency and phase velocity while the wavelength is preserved.
        \textbf{h}, Frequency spectra before ($t<8$~ns, black) and after ($t>12$~ns, red) the temporal interface, revealing the generated negative-frequency component after the field modulation.
        \textbf{i}, Amplitude ratio $|R|/T$ as a function of transition time $\tau$ for different final fields $\mu_0 H_2$. The curves show a rapid suppression of the reflected component with increasing transition time.
        \textbf{j}, Time evolution of $\max(|m_z|)$ for different final fields $\mu_0 H_2$, showing field-dependent amplification or decay after the temporal interface.
        \textbf{k}, FWHM evolution of the wavepacket for different final fields, showing weak post-interface broadening over the analyzed time window.
        \textbf{l}, Total energy density $E_\mathrm{total}$ as a function of time for $\mu_0 H_2=225$ and $230$ mT. Dashed vertical lines mark the temporal-interface window between 10 and 12~ns. The inset shows the post-interface energy change $E(t)-E(15~\mathrm{ns})$ in J/m$^3$, highlighting the slow decrease of the energy density after the switching event.
        } \label{fig:fig_4}
    \end{figure}

    \subsection{Smooth temporal interfaces for frequency conversion and amplification}
 
        Sharp temporal interfaces induce reflections that may be unwanted. To suppress them, we consider smooth interfaces where the field changes over finite time $\tau$.
        By analogy with the Landau-Zener problem in quantum mechanics, which describes two-level systems where the probability of avoiding an energy level transition decays exponentially when parameters change at a finite rate (with the exponent proportional to the square of the energy splitting divided by the rate of change)~\cite{landau1932theorie,zener1932non}, we expect that when the field changes adiabatically (slowly compared to the spin-wave precession frequency) reflections can be suppressed. 
        Specifically, if:
        \begin{equation}
            |\dot{\Omega}|=\frac{\Delta \Omega_0 }{\tau}\ \ll \bar{\Omega}_{0}^2,
            \label{eq:adiabatic_criterion}
        \end{equation}
        where $\bar{\Omega}_{0} = (\Omega_1 + \Omega_2)/2$ and $\Delta\Omega_{0} = |\Omega_2 - \Omega_1|$, the spin wave adiabatically follows the changing dispersion without reflection. We predict:
        \begin{equation}
            \frac{|R|}{T} \sim \exp\left(-C\frac{\bar{\Omega}_{0}^2}{|\dot{\Omega}|}\right),
            \label{eq:reflection_suppression}
        \end{equation}
        where $C$ is determined from the micromagnetic simulations. 
        A full WKB treatment~\cite{Griffiths2018}, along the lines of earlier adiabatic analyses of spin and magnetostatic waves in time-varying magnetic fields~\cite{Rezende1969,Preobrazhenskii1988}, would provide analytically $C$, but it is beyond this work's scope.

        We model smooth temporal interfaces using:
        
        \begin{equation}
            H_y = H_1 - 0.5 \, (H_1 - H_2) \left[ \tanh\left(\frac{4(t - t_i)}{\tau}\right) + 1 \right],
            \label{eq:singleSmoothInterface}
        \end{equation}
        where $t_i=10$~ns is the temporal interface center, $\tau$ is the temporal width, $\mu_0  H_1 = 300$~mT is the initial (high) field, and $\mu_0  H_2$ is the final (low) field value we vary in the range $224$--$260$~mT. At $t = t_i + \tau/2$, approximately 98\% of the transition is complete.
        
        Increasing the transition time $\tau$ suppresses temporal reflections while preserving refracted amplitude (Fig.~\ref{fig:fig_4}a--i); for fields too far below $H_c$ (e.g. $\mu_0 H_2 = 224$~mT) the exponential growth quickly saturates and strong full width at half maximum (FWHM) broadening signals stripe-domain formation (see Supplementary Fig.~S6).

        A combined space--time and spectral analysis confirms that the wavelength (and thus the wavevector) is preserved while the frequency is continuously changed during the smooth temporal ramp, so that temporal refraction generates a strong negative-frequency component at $-0.42$~GHz from an incident packet at $4.09$~GHz (Fig.~\ref{fig:fig_4}g,h).

        Numerical data follow the predicted exponential suppression $|R|/T \sim \exp(-C\bar{\Omega}_0^2/|\dot{\Omega}|)$ with a fitted constant $C \approx 0.5$ (Fig.~\ref{fig:fig_4}i)  validating the Landau-Zener analogy.
        For the transition from $\mu_0 H_1 = 300$ to $\mu_0 H_2 = 225$~mT (evaluated at $k = k_{\mathrm{soft}}$, $f_0=4.09$~GHz, $f_1=-0.42$~GHz), the reflection amplitude decreases to approximately 10\% for $\tau=1$~ns and 1\% for $\tau=2$~ns relative to sharp-interface values.

        Amplitude traces and FWHM evolution show that for $H_2$ below $H_c$ we indeed observe exponential amplitude growth, while for $H_2 > H_c$ there is exponential decay, with rates increasing as we move further from $H_c$. 
        Importantly, the FWHM analysis confirms that within the chosen field range and time window, the wavepacket width remains nearly unchanged, ensuring stable amplification conditions (Fig.~\ref{fig:fig_4}j,k).
        Long-time simulations at $\mu_0 H_2 = 224$, $225$, and $230$~mT, presented in Sec.~S7 of the Supplementary Information, characterize the upper limits of this regime: the linear approximation remains accurate while $\max(|m_z|) \lesssim$ a few percent of $M_s$, beyond which weakly nonlinear broadening and ultimately stripe-domain nucleation set in.
        Thus, the optimal operational window is the slow-instability regime $H_{\mathrm{EP}} < H_2 < H_c$, with both boundaries shifted by DMI as quantified in Methods (Eqs.~\ref{eq:h_c}--\ref{eq:h_EP}): $H_2$ must be  below $H_c$ to provide damping-induced gain, yet remain sufficiently above $H_{\mathrm{EP}}$ so that the growth rate stays controllable and the slow-instability amplification does not collapse into the fast, strong-instability--driven stripe-domain transition.

        To address the apparent paradox of damping-induced gain in a passive system, we examine the total energy density of the simulation volume (Fig.~\ref{fig:fig_4}l), computed directly by MuMax3~\cite{vansteenkiste2014design}.
        The switch from $\mu_0 H_1$ to $\mu_0 H_2$ \emph{raises} $E_\mathrm{total}$ by about $100$~kJ/m$^3$, reflecting the work done by the time-varying field as it reshapes the Zeeman landscape --- the only external energy input.
        For $H_2 < H_c$ the uniformly magnetized state becomes metastable, and after the switch $E_\mathrm{total}$ slowly \emph{decreases} while the spin-wave amplitude grows (inset of Fig.~\ref{fig:fig_4}l, and Fig.~\ref{fig:fig_4}j).
        Such excitations --- spin-wave modes whose creation lowers, rather than raises, the magnetic energy --- have recently been termed \emph{antimagnons}~\cite{Harms2024antimagnonics,Wang2026antimagnonObs}: as the spin-wave amplitude grows, the total magnetic energy of the system decreases, consistently with the strong negative-frequency component at $-0.42$~GHz in the post-interface spectrum (Fig.~\ref{fig:fig_4}h).
        Energy conservation is thus respected: the time-varying field merely prepares a metastable potential, whose excess energy is gradually released into the growing wavepacket of antimagnons.
        The step-like noise in the inset is a numerical artifact of the single-precision recording of $E_\mathrm{total}$ in mumax$^3$, magnified by the small ratio between the spin-wave region and the total simulation volume. A detailed analysis can be found in section S6 of the Supplementary Information.

        \begin{figure}[htbp]
            \centering
            \includegraphics[width=0.95\textwidth]{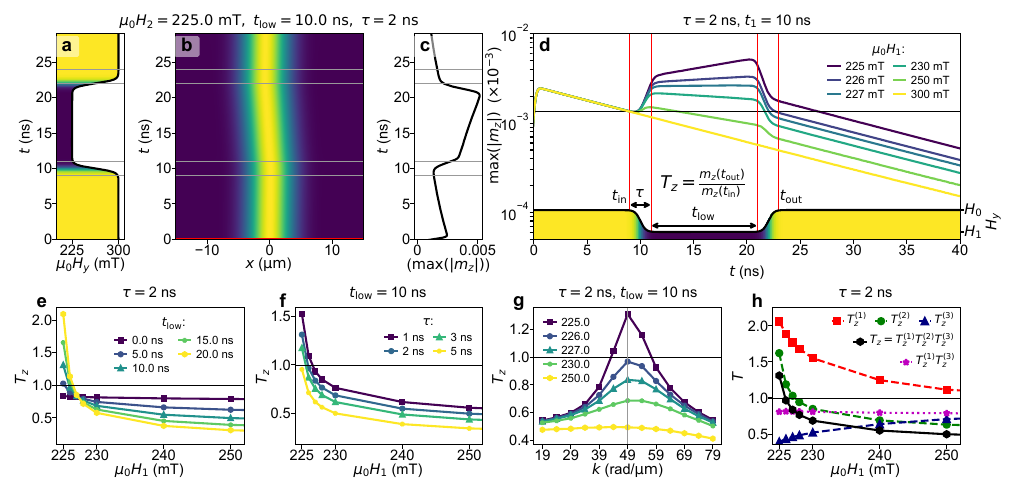}     
            \caption{\textbf{Temporal amplification slab.}
            \textbf{a}, Temporal magnetic field profile showing downward $\mathrm{tanh}$ gradient ($\tau = 2$~ns) from $\mu_0 H_1 = 300$~mT to $\mu_0 H_2 = 225$~mT, constant low-field plateau ($t_\mathrm{low} = 10$~ns), and upward gradient ($\tau = 2$~ns) returning to $\mu_0 H_1$.
            \textbf{b}, Space-time evolution of spin-wave amplitude $m_z(t,x)$ for $\mu_0 H_2 = 225$~mT, normalized to unity at each time step.
            \textbf{c}, Amplitude envelope $\max(|m_z|)$ versus time for the same field value. Horizontal lines indicate temporal landmarks: $t_\mathrm{in} = t_i - \tau/2$ (slab entrance at 98\% field descent) and $t_\mathrm{out} = t_i + t_\mathrm{low} + 1.5\tau$ (slab exit at 98\% field recovery).
            \textbf{d}, Amplitude evolution in logarithmic scale for different low-field values $\mu_0 H_2 = 225$--$300$~mT ($\tau = 2$~ns, $t_\mathrm{low} = 10$~ns, $k_\mathrm{soft}$). Three temporal regions are marked: initial high-field propagation, low-field plateau, and final high-field recovery. Gray horizontal lines (a--c) and red vertical lines (d) mark the temporal interfaces.
            \textbf{e}, Transmission coefficient $T_z$ versus $\mu_0 H_2$ for low-field plateau durations $t_\mathrm{low} = 0, 5, 10, 15, 20$~ns ($\tau = 2$~ns, $k_\mathrm{soft}$).
            \textbf{f}, Transmission coefficient $T_z$ versus $\mu_0 H_2$ for transition time duration $\tau = 1, 2, 3, 5$~ns ($t_\mathrm{low} = 10$~ns, $k_\mathrm{soft}$).
            \textbf{g}, Transmission coefficient $T_z$ versus spin-wave wavevector $k$ for field values $\mu_0 H_2 = 225$--$250$~mT ($\tau = 2$~ns, $t_\mathrm{low} = 10$~ns).
            \textbf{h}, Decomposition of transmission into components: $T_z^{(1)}$ (first interface), $T_z^{(2)}$ (low-field plateau), $T_z^{(3)}$ (second interface), with products $T_z^{(1)} T_z^{(3)}$ and total transmission $T_z = T_z^{(1)} T_z^{(2)} T_z^{(3)}$ versus $\mu_0 H_2$ ($\tau = 2$~ns, $t_\mathrm{low} = 10$~ns, $k_\mathrm{soft}$).
            }\label{fig:fig_5}
        \end{figure}

    \subsection{Frequency-preserving amplification via temporal slabs}
    
        The temporal interface phenomenon enables frequency conversion with amplification. However, many magnonic devices require amplification \emph{without} frequency change to maintain signal integrity.
        Here, we demonstrate that cascading two complementary temporal interfaces---forming a temporal amplification slab---enables net gain while preserving the original frequency. The temporal field profile is described by:
        \begin{equation}
            \begin{aligned}
                H_y = H_1 - &\frac{1}{2}(H_1 - H_2) \Bigl[ 
                    \tanh\left(\frac{4(t - t_i)}{\tau}\right) \\
                    -\, &\tanh\left(\frac{4(t - (t_i + 2\tau + t_{\mathrm{low}}))}{\tau}\right) \Bigr],
            \end{aligned}
            \label{eq:temporal_slab}
        \end{equation}
        where the first $\tanh$ term describes the downward field transition and the second $\tanh$ term describes the upward recovery, separated by the low-field plateau duration $t_{\mathrm{low}}$.
        A spin-wave packet thus experiences a high-field $\to$ low-field $\to$ high-field sequence defined by Eq.~\eqref{eq:temporal_slab}: it enters the slab at $H_1$, dwells at $H_2$ for a time $t_\mathrm{low}$, and exits back at $H_1$, so that the temporal slab can be viewed as a frequency-preserving amplifier acting on a selected $k$--mode (Fig.~\ref{fig:fig_5}a--c).

        Fig.~\ref{fig:fig_5}d quantifies field-dependent behavior across $\mu_0 H_2 = 225$--$300$~mT ($\tau = 2$~ns, $t_\mathrm{low} = 10$~ns, $k_\mathrm{soft}$).
        Three distinct dynamical regimes are visible: (1) downward transition (falling field edge)---all trajectories show the amplitude increase during this interface crossing; (2) constant low-field plateau---amplitude exhibits the exponential behavior determined by $H_2$, namely: exponential growth for $H_2 < H_c$ (approximately $227.2$~mT), exponential decay for $H_2 > H_c$, minimal change at $H_2 \approx H_c$; (3) upward transition (rising external magnetic field)---amplitude decreases as the field returns to $\mu_0 H_1$. The logarithmic scale reveals the exponential characters of growth and decay in the constant-field regions, enabling clear quantification of the amplification rate. 
        Net amplification results from exponential growth on the low-field plateau overcoming the net interface effect, with a modest gain at the entrance and a stronger loss at the exit.
        We define the transmission coefficient of the temporal slab as the ratio of $m_z$ amplitudes at the slab exit and entrance:
        \begin{equation}
            T_z = \frac{\max(|m_z|(t_\mathrm{out}))}{\max(|m_z|(t_\mathrm{in}))},
            \label{eq:transmission}
        \end{equation}
        where $t_\mathrm{in} = t_i - \tau/2$ and $t_\mathrm{out} = t_i + t_\mathrm{low} + 1.5\tau$ denote the temporal positions of the slab beginning and end. At both times, the magnetic field is approximately equal to $H_1$, differing by only $\Delta H_\mathrm{offset} \approx 0.018(H_1 - H_2)$ (approximately 1.8\% of the total field change), so that the precession ellipticity (and hence the magnonic temporal impedance, $Z = \varepsilon^z$) at the entrance and at the exit are essentially identical, $Z(t_\mathrm{in}) \approx Z(t_\mathrm{out}) \approx Z_1$. With this symmetric choice, $T_z$ as defined in Eq.~\eqref{eq:transmission} measures the genuine net change of the $m_z$ amplitude across the slab in a state of equivalent magnonic impedance (including damping accumulated during the finite-duration ramps and plateau), with $T_z > 1$ indicating net amplification and $T_z < 1$ indicating net attenuation.
        To systematically optimize the temporal slab for amplification, we examine the transmission coefficient $T$ as a function of the slab parameters. Figs.~\ref{fig:fig_5}e-g present detailed parametric studies: low-field plateau duration $t_\mathrm{low}$, the transition time $\tau$, and values of spin-wave wavevector $k$, revealing how each parameter influences the amplification characteristics near the critical magnetic field.

        Fig.~\ref{fig:fig_5}e reveals a critical field-dependent transition. For $H_2 > H_c$, increasing $t_\mathrm{low}$ decreases transmission (damping dominance). For subcritical fields ($H_2 < H_c$), the opposite occurs: at $H_2 = 225$~mT, transmission increases from ~1.3 ($t_\mathrm{low} = 10$~ns) to ~2.1 ($t_\mathrm{low} = 20$~ns), while at $H_2 = 226$~mT, gain emerges only at $t_\mathrm{low} \geq 20$~ns. This demonstrates that extended low-field residence in slow instability enables amplification.
        
        Fig.~\ref{fig:fig_5}f shows how transmission decreases monotonically with increasing $\tau$. Although $\tau = 1$~ns yields highest transmission numerically, Fig.~\ref{fig:fig_4}i reveals that sharp transitions produce reflections (~10\% $|R|/T$ per interface) with the interference oscillations degrading wavepacket quality. At $\tau = 2$~ns, transmission remains near-optimal while eliminating these artifacts. A further $\tau$ increase to $5$~ns reduces transmission as the damping accumulation at fields above $H_c$ outweighs adiabatic benefits.
        
        Fig.~\ref{fig:fig_5}g reveals wide peaks in transmission versus $k$ whose width remains constant but whose amplitudes increase dramatically as the field decreases toward $H_c$. The maximum peak height occurs near $k_\mathrm{soft}$ at $\mu_0 H_2 = 225$~mT, while peaks nearly vanish for fields higher than $H_c$. This wavenumber selectivity arises from matching to the softened spin-wave mode whose frequency approaches zero at the critical field.

        Fig.~\ref{fig:fig_5}h decomposes the transmission coefficient into three multiplicative contributions corresponding to the three stages of the slab:
        \begin{equation}
            T_z = T_z^{(1)} \, T_z^{(2)} \, T_z^{(3)},
            \label{eq:Tz_decomposition}
        \end{equation}
        
        where $T_z^{(1)}$ describes the impedance transformation $Z_1 \to Z_2$ at the entrance interface, $T_z^{(2)}$ captures the dynamical evolution of the wavepacket on the low-field plateau (medium~2, with magnonic impedance $Z_2$), and $T_z^{(3)}$ describes the symmetric impedance transformation $Z_2 \to Z_1$ at the exit interface.
        The critical observation is that $T_z^{(1)} T_z^{(3)} < 1$ across all fields: the two interfaces alone act as passive impedance transformers---temporal analogues of an antireflection coating realized as a smooth impedance taper---and net-attenuate the wavepacket.
        At high fields, there is no net gain, $T_z^{(2)} < 1$.
        For $H_2<H_c$, $T_z^{(2)}$ grows above 1 (reaching $\sim 2$ at $\mu_0 H_2 = 225$~mT), offsetting the interface losses; all useful gain therefore resides in $T_z^{(2)}$, i.e. in the slow-instability dynamics on the low-field plateau.

        These results lead to several design conclusions.
        Net amplification requires operation in the slow-instability window $H_{\mathrm{EP}} < H_2 < H_c$: all useful gain resides in $T_z^{(2)}$ (slow instability with $\mathrm{Im}\,\Omega_+ > 0$). For lower fields the dynamics enter the strong-instability regime, where rapid stripe-domain nucleation within the $t_\mathrm{low}$ plateau degrades the wavepacket. For our CoFeB stack this operational trade-off is met for $\mu_0 H_2 \approx 225$--227~mT.
        Moreover, since $T_z^{(1)}T_z^{(3)} < 1$ for all fields (the interfaces act as passive impedance transformers), net gain requires $T_z^{(2)} \gg 1$, i.e., sufficient dwell time in the slow-instability regime to offset interface attenuation.
        Finally, optimization therefore prioritizes: (a) biasing $H_2$ within the slow-instability window; (b) adiabatic transitions ($\tau \approx 2$~ns) that suppress temporal reflections and stabilize $T_z^{(1)}$; (c) low-field plateaus long enough to build up $T_z^{(2)}$; and (d) wavevector selection around the softened-mode.

\begin{figure}[htbp]
            \centering
            \includegraphics[width=0.95\textwidth]{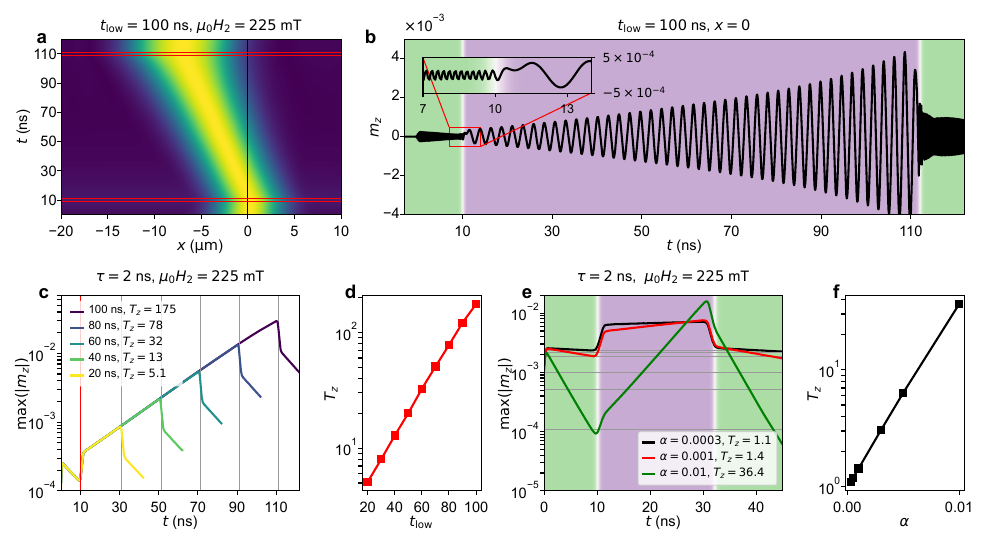}     
            \caption{\textbf{Giant temporal amplification and its dependence on damping.}
            \textbf{a}, Space-time evolution of the spin-wave envelope (normalized to unity at each time step) through an optimized temporal slab ($t_\mathrm{low} = 100$~ns, $\mu_0 H_2 = 225$~mT, $\tau = 2$~ns). 
            \textbf{b}, Time evolution of $m_z$ at fixed position ($x = 0$), showing oscillation growth across the temporal slab. Inset: smooth frequency transition at the temporal interface. Background colors mark temporal regions: green (high field, $\mu_0 H = 300$~mT), purple (low-field plateau, $\mu_0 H = 225$~mT), with smooth transitions indicating adiabatic field changes ($\tau = 2$~ns).
            \textbf{c}, The maximum amplitude $\max(|m_z|)$ versus time for five plateau durations ($t_\mathrm{low} = 20, 40, 60, 80, 100$~ns). The legend shows exponential growth rates increase with $t_\mathrm{low}$, demonstrating cumulative amplification with extended slab residence time.
            \textbf{d}, Transmission coefficient $T_z$ against the plateau duration $t_\mathrm{low}$, showing the exponential scaling $T_z \propto \exp(\mathrm{const}\cdot t_\mathrm{low})$, reaching $175$-fold amplification at $t_\mathrm{low} = 100$~ns.
            \textbf{e}, The amplitude evolution for three different damping values ($\alpha = 0.0003, 0.001, 0.01$) at fixed slab parameters ($t_\mathrm{low} = 20$~ns, $\tau = 2$~ns, $\mu_0 H = 225$~mT). Larger damping speeds growth in the low-field plateau and decay outside, counterintuitively increasing net amplification.
            \textbf{f}, Transmission coefficient versus damping parameter, revealing exponential dependence $T_z \propto \exp(\Omega_i t_\mathrm{low})$ with $\Omega_i \propto \alpha$. The straight line represents fit according to Eq.~\eqref{eq:Im_unified} and field value approximately 2~mT below $\mu_0 H_c$.
            }\label{fig:fig_6}
        \end{figure}

    \subsection{Giant amplification via temporal slabs}

        We examine the ultimate performance limits of optimized temporal slabs under near-optimal conditions: field within slow instability regime ($\mu_0 H_2 = 225$~mT), extended low-field plateau durations ($t_\mathrm{low} = 20$--$100$~ns), smooth temporal transitions ($\tau = 2$~ns) and $k = k_{\mathrm{soft}}$.
        The input wavepacket amplitude is set to $\max(|m_z|) \sim 10^{-4} M_s$, representative of small-amplitude excitations driven by standard microwave antennas at experimentally accessible signal-to-noise ratios.

        Figs.~\ref{fig:fig_6}a--d demonstrate the resulting amplification: $T_z$ reaches 5-, 13-, 32-, 78-, and 175-fold for $t_\mathrm{low} = 20, 40, 60, 80,$ and $100$~ns, respectively, growing exponentially with the plateau duration (Fig.~\ref{fig:fig_6}d).
        Fig.~\ref{fig:fig_6}b confirms the smooth, coherent amplitude evolution of the wavepacket throughout the slab.
        We emphasize that the 175-fold value is not an intrinsic ceiling of the mechanism but rather an operational compromise: at $t_\mathrm{low} = 100$~ns the output amplitude reaches $\max(|m_z|) \sim 2\%$ of $M_s$, which approaches the upper edge of the linear regime in which the wavepacket character is preserved (cf. Fig.~\ref{fig:fig_4}j,k and Sec.~S7 of the Supplementary Information).
        A smaller input amplitude or longer $t_\mathrm{low}$ would yield a proportionally larger numerical gain, but at the cost of either reduced experimental signal-to-noise ratio at the input or nonlinear saturation and stripe-domain nucleation at the output.
        The reported 175-fold value should therefore be read as a representative gain achievable under realistic operating conditions, rather than as an absolute upper bound.
        
        Remarkably, transmission exhibits the same exponential structure when varying damping rather than plateau duration. A systematic comparison across damping values ($\alpha = 0.0003, 0.001, 0.01$) at fixed temporal parameters ($\tau = 2$~ns, $t_\mathrm{low} = 20$~ns) shows that larger damping produces stronger amplification (Figs.~\ref{fig:fig_6}e,f), with $T_z \propto \exp(\mathrm{const}\cdot \alpha)$. This counterintuitive behavior---where increased dissipation enhances gain---is fully consistent with the analytical model: in the slow-instability regime, $\mathrm{Im}(\Omega_+)$ scales linearly with $\alpha$ (Eq.~\eqref{eq:Im_unified}). Since amplitude evolves as $|m| \propto \exp[\mathrm{Im}(\Omega) t]$, the transmission follows $T_z \propto \exp(\mathrm{const} \cdot \alpha \cdot t_\mathrm{low})$, explaining why both $t_\mathrm{low}$ and $\alpha$ contribute to the exponential scaling through the same functional form.

\begin{figure}[htbp]
        \centering
        \includegraphics[width=0.95\textwidth]{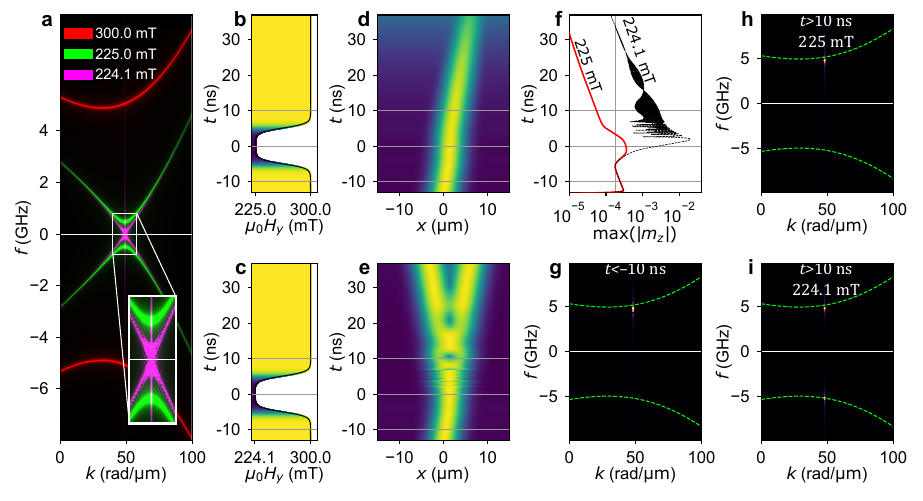}     
        \caption{\textbf{Temporal slab in a PMA film without DMI.}
        \textbf{a}, Spin-wave dispersion for a CoFeB film with perpendicular magnetic anisotropy and no DMI ($D = 0$; all other parameters as in the main text), shown for three applied fields: $300$~mT (red), $225$~mT (green) and $224.1$~mT (magenta). The dispersion remains mirror-symmetric about $k = 0$ for all fields.
        \textbf{b}, \textbf{c}, Temporal profiles of the applied field in a temporal-slab protocol with smooth ramps of duration $\tau = 5$~ns and plateau time $t_\mathrm{low} = 5.1$~ns, where the field is reduced from $300$~mT to $225$~mT (\textbf{b}) or $224.1$~mT (\textbf{c}) and subsequently restored to $300$~mT.
        \textbf{d}, \textbf{e}, Spatial profiles of the spin-wave amplitude $|m_z(x,t)|$ during the descending ramp for the two target fields. For $224.1$~mT (\textbf{e}) a reflected wave packet is visible at the temporal interface.
        \textbf{f}, Time evolution of the maximum spin-wave amplitude for temporal slabs with target fields $225$~mT (solid red) and $224.1$~mT (dashed black).
        \textbf{g}, Spatio-temporal Fourier spectrum (2D FFT in space and time) for $t < 10$~ns, showing the incident spin-wave packet with well-defined wavelength and frequency.
        \textbf{h}, \textbf{i}, Spatio-temporal Fourier spectra for $t > 10$~ns for the two target fields, corresponding to the temporal slabs in \textbf{b} and \textbf{c}, respectively.
        }
        \label{fig:noDMI}
    \end{figure}

\subsection{Strong-instability amplification in PMA-only systems}

    Interfacial DMI plays a dual role in the slow-instability mechanism: it separates $H_\mathrm{EP}$ from $H_c$ to open the operational window $\Delta H \propto D^2/M_s^3$, and it breaks mirror symmetry so that only one dispersion branch softens to $\omega \to 0$, enabling adiabatic tracking with exponentially suppressed reflections.
    A natural question, of particular relevance for the broad class of ferromagnetic thin films that exhibit PMA without interfacial DMI, is whether phase-transition-driven amplification persists once DMI is removed.

    To answer this, we simulate a PMA-only film ($D = 0$) with otherwise identical parameters (Fig.~\ref{fig:noDMI}). The dispersion remains mirror-symmetric about $k = 0$ for all fields (Fig.~\ref{fig:noDMI}a), and $H_c$ coincides with $H_\mathrm{EP}$. When the temporal slab reduces the field to $\mu_0 H_2 = 225$~mT, which lies above $H_c$, the wavepacket tracks adiabatically with negligible reflection and the post-interface spectrum shows a single positive-frequency peak (Fig.~\ref{fig:noDMI}d,h)---consistent with the damping regime where $\mathrm{Im}(\Omega_+) < 0$.

    The situation changes when the field is driven just below $H_c$, e.g.\ to $\mu_0 H_2 = 224.1$~mT.
    The spatial profiles now display a pronounced reflected wavepacket with the same amplitude the refracted one (Fig.~\ref{fig:noDMI}e), and the post-interface spectrum develops two peaks at the same $k$ but frequencies with opposite signs (Fig.~\ref{fig:noDMI}i).
    The downward field ramp carries the mode along its dispersion branch until the frequency reaches $f = 0$ at $H_c = H_\mathrm{EP}$, where both branches coalesce; during the upward ramp, the excitation splits symmetrically onto the two branches, redistributing energy equally between forward- and backward-propagating wavepackets.

    The strong-instability regime in PMA-only systems thus provides an alternative amplification route that does not require interfacial DMI and is therefore accessible in a substantially wider class of materials.
    The trade-off is that the gain is intrinsically accompanied by strong back-reflection and rapid stripe-domain nucleation, and is therefore harder to control than the slow-instability mechanism with DMI.
    DMI is thus not strictly required for phase-transition-driven instability, but it is essential for (i)~opening a finite \emph{slow}-instability window with $\alpha$-linear gain above the strong instability, and (ii)~suppressing reflection losses at temporal interfaces.    

\section{Discussion}

Our results demonstrate that ultrathin ferromagnetic films with perpendicular anisotropy and interfacial DMI exhibit damping-induced spin-wave amplification when subjected to temporal magnetic field modulation near the stripe-domain phase transition.


From an application perspective, our results show that a spatially uniform film with PMA and interfacial DMI, subject to a time-modulated field, can operate as a high-gain, frequency-preserving spin-wave amplifier without any lithographic patterning or continuous microwave drive.
At a single temporal interface, any impedance mismatch $Z_1 \neq Z_2$ expands the precession ellipse and, for fields between $H_\mathrm{EP}$ and $H_c$, creates the conditions for slow-instability-driven gain.
The temporal slab amplifier composes two such impedance transformations with an intervening propagation in medium~2: the entrance interface ($T_z^{(1)}$, $Z_1 \to Z_2$) and the exit interface ($T_z^{(3)}$, $Z_2 \to Z_1$) act as passive impedance transformers with $T_z^{(1)} T_z^{(3)} < 1$, while all useful gain is accumulated in $T_z^{(2)}$ during propagation in the slow-instability regime.


The damping-induced gain mechanism does not require exotic ultra-low-damping materials.
Standard CoFeB films with $\alpha \sim 10^{-2}$ produce excellent amplification, since the growth rate scales linearly with $\alpha$ in the slow-instability regime.
Materials with even higher damping---such as Co/Pt and Co/Pd multilayers~\cite{Lesniewski2025}---are correspondingly favourable, provided they exhibit the required field-driven phase transition and a symmetry-breaking mechanism that opens the slow-instability window.

The analytical model developed here provides explicit expressions for the complex frequency in each dynamical regime and yields three key design rules: (i) the amplification window width $\Delta H \propto D^2/M_s^3$ is independent of damping, (ii) the growth rate in the slow instability regime scales linearly with $\alpha$, and (iii) the ratio of growth rates between slow and strong instability is $\sim \alpha \omega_D/(2\omega_x) \approx 0.01$--$0.1$ for typical parameters. These scaling laws guide material selection and operating point optimization.

The mechanism exhibits intrinsic wavevector selectivity, peaking around $k_{\mathrm{soft}} \approx 49$~rad/\textmu m for the CoFeB film analyzed here, with an effective bandwidth that can exceed $20$~rad/\textmu m for fields approaching $H_\mathrm{EP}$ (Figs.~\ref{fig:fig_5}g and~\ref{fig:k_selectivity}d).
This $k$-selectivity provides a built-in filtering mechanism for application-specific bandwidth and can be tuned in other material platforms by engineering $k_\mathrm{soft}$.
Critically, reversing the temporal sequence reverses the amplification, enabling on-demand control without any permanent device modification.

As demonstrated in Fig.~\ref{fig:noDMI}, PMA-only systems ($D = 0$) can also exhibit amplification, but with important limitations. Without DMI, the critical field coincides with the exceptional point ($H_c = H_{\mathrm{EP}}$), eliminating the slow-instability window entirely. Consequently, any field excursion below $H_c$ immediately enters the strong-instability regime, where growth rates are higher ($\propto \sqrt{|\omega_x|\omega_z}$) but stripe-domain nucleation is more difficult to control. Furthermore, the symmetric dispersion causes both branches to coalesce at $f = 0$ during adiabatic field ramps, producing unavoidable back-reflections even for smooth temporal interfaces (Fig.~\ref{fig:noDMI}e,i). This trade-off---faster amplification at the cost of time-reflected spin waves  and reduced controllability---may nevertheless prove advantageous in applications where maximum gain per unit time is prioritized over wavepacket fidelity. 
This trade-off---faster amplification at the cost of reflection losses and reduced controllability---positions PMA-only systems as a viable alternative within a broader material class, particularly for applications where maximum gain per unit time is prioritized over wavepacket fidelity.

The framework developed here is not specific to CoFeB but applies generally to any ferromagnetic thin film that exhibits (i) a field-driven phase transition between uniform magnetization and a modulated state (e.g., stripe domains), and (ii) sufficient DMI or other symmetry-breaking mechanism to open the slow-instability window.
Although our analytical model is formulated for interfacial DMI, other mechanisms that break mirror symmetry of the spin-wave dispersion should enable analogous behavior. These include asymmetric multilayer stacks, e.g., NdCo/Al/Py~\cite{Szulc2022}, graded material-parameter profiles along the film thickness~\cite{gallardo2019, christienne2025}, or placing a perfect electric conductor~\cite{mruczkiewicz2014} or superconductor~\cite{golovchanskiy2018} on one of the film surfaces. The key requirement is that only one dispersion branch softens toward $\omega \to 0$ at finite $k$ near the phase transition, while the counter-propagating branch remains at finite frequency---enabling adiabatic tracking and reflection-free temporal interfaces. We expect that the width of the slow-instability window would scale with the strength of the symmetry-breaking mechanism, analogously to the $D^2/M_s^3$ scaling derived here for DMI. Experimental verification of temporal amplification using these alternative nonreciprocity sources remains an open direction.

Both PMA and interfacial DMI are experimentally tunable in ferromagnet/nonmagnetic heavy metal heterostructures PMA through interface composition, layer thickness, and voltage-controlled anisotropy~\cite{Soumyanarayanan2017,yamamoto2022vcma}, and interfacial DMI through asymmetric heavy-metal/ferromagnet interfaces, additive chiral interactions in multilayers, and garnet-based heterostructures
~\cite{MoreauLuchaire2016,Soumyanarayanan2017,Ding2019,fakhrul2024}.
Beyond ferromagnet/nonmagnetic heavy metal heterostructures, iron-garnet thin films with PMA provide a complementary platform with a fundamental advantage rooted in their material parameters: the much smaller saturation magnetization yields a substantially larger exchange length $l_\mathrm{ex}$, which through Eq.~\eqref{eq:k_soft} translates into an order-of-magnitude smaller $k_\mathrm{soft}$ and, correspondingly, several times longer operating wavelength. PMA has been demonstrated in Ce:YIG~\cite{ghising2017}, Bi-substituted YIG~\cite{das2024}, and TmIG~\cite{prestwood2025}, and interfacial DMI has been reported at garnet/substrate and garnet/heavy-metal interfaces~\cite{wang2020,Ding2019,fakhrul2024}. As detailed in Table~\ref{tab:SM_materials}, TmIG-based systems exhibit $\lambda_\mathrm{soft} \approx 0.7$--$0.9$~\textmu m, compared with $\lambda_\mathrm{soft} \approx 130$~nm for the CoFeB film analyzed here---bringing the operating wavelength into the range of standard optical techniques such as Brillouin light scattering and time-resolved MOKE microscopy. The trade-off is a substantially narrower slow-instability window ($\Delta H$ ranging from $\sim 30$~\textmu T to $\sim 0.3$~mT for representative TmIG parameters, cf.\ Table~\ref{tab:SM_materials}), which sets a more demanding requirement on field-control precision, although the underlying amplification rate per unit time remains comparable.

\begin{table}
    \centering
    \renewcommand{\arraystretch}{0.85}
    \caption{Material parameters and estimated slow-instability characteristics for CoFeB and iron garnet systems with interfacial DMI. Material parameters for TmIG-based systems are taken from the cited references; values in parentheses are estimates based on typical garnet properties. Derived quantities are calculated using the equations referenced in the first column.}
    \label{tab:SM_materials}
    \footnotesize
    \begin{tabular}{@{}l c c c c@{}}
    \toprule
    \textbf{Parameter} & \textbf{Unit} & \textbf{CoFeB} & \textbf{TmIG/Pt} & \textbf{TmIG/BiYIG} \\
     & & (this work) & \cite{Ding2019} & \cite{fakhrul2024} \\
    \midrule
    \multicolumn{5}{c}{\textit{Material parameters}} \\
    \midrule
    $d$ & nm & 2 & 15 & 30\textsuperscript{a} \\
    $M_s$ & kA/m & 1420 & 110 & 90\textsuperscript{b} \\
    $A_{\mathrm{ex}}$ & pJ/m & 13 & (4.0)\textsuperscript{c} & (4.0)\textsuperscript{c} \\
    $D$ & mJ/m$^2$ & 0.5000 & 0.0150 & 0.0145 \\
    $Q$ & --- & 1.1 & (1.0)\textsuperscript{d} & (0.9)\textsuperscript{e} \\
    $\alpha$ & --- & 0.002 & 0.015 & (0.006)\textsuperscript{f} \\
    \midrule
    \multicolumn{5}{c}{\textit{Derived parameters}} \\
    \midrule
    $l_{\mathrm{ex}}= \sqrt{A_{\mathrm{ex}}/(\tfrac{1}{2}\mu_0 M_s^2)}$ & nm & 3.2 & 22.9 & 28.0 \\
    $l_D= D/(\tfrac{1}{2}\mu_0 M_s^2)$ & nm & 0.39 & 1.97 & 2.85 \\
    \midrule
    \multicolumn{5}{c}{\textit{Soft-mode characteristics}} \\
    \midrule
    $k_{\mathrm{soft}}= d/(4 l_{\mathrm{ex}}^2)$, Eq.~\eqref{eq:k_soft} & rad/\textmu m & 48.6 & 7.1 & 9.5 \\
    $\lambda_{\mathrm{soft}}=2\pi/k_\mathrm{soft}$ & nm & 130 & 880 & 660 \\
    \midrule
    \multicolumn{5}{c}{\textit{Field boundaries and amplification window}} \\
    \midrule
    $\mu_0 H_{\mathrm{EP}}$, Eq.~\eqref{eq:h_EP} & mT & 219.3 & 3.4 & 232.7 \\
    $\mu_0 H_c$, Eq.~\eqref{eq:h_c} & mT & 222.5 & 3.7 & 232.8 \\
    $\Delta H$, Eq.~\eqref{eq:Delta_h} & mT & 3.20 & 0.27 & 0.037 \\
    \midrule
    \multicolumn{5}{c}{\textit{Amplification rate at midpoint}} \\
    \midrule
    $\Gamma_{\mathrm{mid}}/2\pi$ at $H\!=\!(H_c\!+\!H_{\mathrm{EP}})/2$, Eq.~\eqref{eq:Im_unified} & MHz & 4.08 & 1.26 & 1.50 \\
    $\tau = 1/\Gamma_{\mathrm{mid}}$ & ns & 39 & 127 & 106 \\
    \bottomrule
    \end{tabular}
    
    \vspace{1ex}
    \raggedright\footnotesize
    \textsuperscript{a}~Hypothetical thickness chosen to illustrate scaling; experimental samples in Ref.~\cite{fakhrul2024} have $d = 6.3$~nm.\\
    \textsuperscript{b}~Mid-range value; Ref.~\cite{fakhrul2024} reports $M_s = 80$--$105$~kA/m for BiYIG/TmIG bilayers.\\
    \textsuperscript{c}~Assumed equal to bulk YIG value; $A_{\mathrm{ex}}$ not reported in Refs.~\cite{Ding2019,fakhrul2024}.\\
    \textsuperscript{d}~Estimated from typical TmIG parameters.\\
    \textsuperscript{e}~Calculated from $K_u = 4.6$~kJ/m$^3$ and $M_s = 90$~kA/m reported in Ref.~\cite{fakhrul2024}.\\
    \textsuperscript{f}~Estimated from BiYIG/TmIG bilayer damping $\alpha = 5$--$7.6 \times 10^{-3}$ in Ref.~\cite{fakhrul2024}; pure BiYIG has $\alpha \approx 5 \times 10^{-4}$.
\end{table}

From an experimental standpoint, a key question is whether the bias field must be modulated uniformly across the entire sample or only locally. Since our mechanism relies on wavepacket propagation through a temporal interface, it suffices to vary the field amplitude only in the region where the wavepacket is localized at a given instant. This significantly relaxes experimental requirements: even a microstrip antenna inducing a spatially localized field modulation should work effectively, provided the temporal profile is synchronized with the wavepacket arrival.

Our demonstration employs electromagnet modulation, while device-level implementations could leverage voltage-controlled magnetic anisotropy (VCMA) for sub-nanosecond switching and direct $H_c$ control~\cite{yamamoto2022vcma}. In our ultrathin CoFeB film with PMA and interfacial DMI, $k_{\mathrm{soft}}$ is pinned to the dispersion minimum at $H = H_c$, where $f = 0$ and $v_g(k_{\mathrm{soft}}) = 0$, so amplification in the slow-instability window $H_{\mathrm{EP}} < H < H_c$ targets modes in a low--$v_g$ sector of the spectrum, whereas efficient routing outside the temporal slab benefits from larger group velocities. This motivates co-engineering the dispersion landscape and temporal control so that the mode traversing the temporal slab connects to high--$v_g$ states in the surrounding propagation regions, for example via VCMA-driven tuning of PMA or other temporal control of PMA, DMI, saturation magnetization, or exchange stiffness, thereby enabling temporal amplification of spin waves that propagate rapidly outside the slab.

Beyond amplification, the efficient frequency conversion (range from 4.09~GHz to $-0.42$~GHz, Fig.~\ref{fig:fig_4}h) suggests applications in magnonic frequency mixing and signal processing.
Furthermore, the unified description of damping and slow instability regimes reveals a remarkable feature at the critical field $H_c$: both $\mathrm{Re}(\Omega)$ and $\mathrm{Im}(\Omega)$ vanish simultaneously. This coincidence is not accidental---it follows directly from Eq.~\eqref{eq:Im_unified}, where $\omega_x = \omega_x^*$ yields $\mathrm{Im}(\Omega) = 0$ at exactly the same field where $\mathrm{Re}(\Omega) = \sqrt{\omega_x\omega_z} - \omega_D = 0$. At this marginal stability point, spin waves are effectively ``frozen'': they neither oscillate nor decay. Combined with the vanishing group velocity $v_g \to 0$ at the dispersion minimum, this creates a condition for spin-wave storage. Unlike photonic slow-light systems\cite{liu2001, fleischhauer2005, hau1999}, where absorption typically limits storage time (since $\mathrm{Im}(\omega) < 0$ persists even as $v_g \to 0$), our system offers a fundamentally different regime where both propagation and dissipation are simultaneously suppressed. By operating slightly below $H_c$, one can even achieve slow-instability conditions with net amplification ($\mathrm{Im}(\Omega) > 0$), compensating for residual losses. While photonic time-varying media have demonstrated related capabilities through different mechanisms, the magnonic realization offers complementary advantages: field-driven reversibility and intrinsic coupling to magnetic phase transitions.

Several limitations of this study warrant discussion.
First, our predictions rely on micromagnetic simulations; experimental validation remains essential. As discussed in the section on PMA-only systems above, two complementary pathways exist: systems with nonreciprocal dispersion (e.g., CoFeB/Pt with interfacial DMI) enable clean amplification with smooth temporal interfaces, whereas reciprocal systems exhibit both refracted and reflected components.

Second, the experimental detection scheme is platform-specific.
For the CoFeB system analyzed here ($\lambda_{\mathrm{soft}} \approx 130$~nm), two complementary routes are available: (i)~all-electrical \emph{propagating spin-wave spectroscopy} using coplanar-waveguide antennas placed on either side of the temporal slab~\cite{yu2013,che2020}, and (ii)~\emph{X-ray magnetic microscopy} (e.g.\ STXM)~\cite{grafe2019,traeger2021}. The longer operating wavelengths of iron-garnet platforms (Table~\ref{tab:SM_materials}) bring them within the range of standard optical techniques such as micro-focused Brillouin light scattering and super-Nyquist sampling time-resolved MOKE microscopy. For platforms in which the slow-instability window is impractically narrow, the strong-instability regime ($H < H_\mathrm{EP}$) provides an alternative route accessible to PMA-only iron garnets providing longer operating wavelengths~\cite{Lesniewski2025}.

Third, the narrow operational window ($\Delta H \approx 3$--$4$~mT for CoFeB; cf.\ Table~\ref{tab:SM_materials} for other platforms) demands precise field control at the few-mT level, achievable with modern electromagnets. The window can be widened by stronger DMI or alternative symmetry-breaking mechanisms.

Fourth, the analytical model relies on linearized Landau--Lifshitz--Gilbert dynamics~\cite{GurevichMelkov1996}, valid in the small-amplitude regime $|m_x|, |m_z| \ll M_s$. Once the precession amplitude becomes comparable to $M_s$, nonlinear dynamics of spin waves break the linearization and the wavepacket character is lost through stripe-domain nucleation~\cite{Kisielewski2023} (cf.\ Sec.~S7 of the Supplementary Information). The model also adopts the uniform mode through the thickness, thickness-averaged dipolar approximation, which introduces a few-percent shift in the predicted $H_c$ and $H_{\mathrm{EP}}$ near $k_{\mathrm{soft}}$ relative to full micromagnetic results.
For thicker films hosting stripe-domain patterns, analogous effects are expected, but their theoretical study would require either full micromagnetic simulations, direct numerical integration of the Landau--Lifshitz--Gilbert equation, or an extended analytical model that relaxes these approximations.

Fifth, our simulations neglect thermal noise and structural defects. The signal-to-noise ratio sets a lower bound on the amplifiable wavepacket amplitude, and defects may both scatter propagating spin waves and act as spin-wave sources under abrupt field changes~\cite{davies2015}.

To our knowledge, this work provides the first systematic analytical and computational framework for spin-wave scattering at temporal magnetic interfaces that incorporates dipolar, exchange, PMA, and DMI contributions. The demonstrated 175-fold amplitude amplification through reversible, lithography-free field modulation compares favorably with existing parametric and spin-torque schemes, which typically achieve gains of 10--50-fold while requiring continuous power injection or fixed structural elements (see table S2 in Supplemental Information for comparison of different amplification schemes). Looking forward, periodic temporal modulation of the magnetic field could give rise to magnonic analogues of photonic time crystals and momentum-space band gaps, while cascaded temporal slabs may enable higher cumulative gain. The ability to exploit field-driven magnetic phase transitions---offering rapid, reversible, and precisely controllable dispersion changes---combined with the synergy between damping and amplification identified here, suggests that temporal field modulation may complement spatial structuring as a practical tool for reconfigurable magnonic signal processing.

This work reveals two key properties of ultrathin magnetic films with perpendicular anisotropy and interfacial DMI subjected to temporal field modulation.

First, spin-wave scattering at temporal magnetic interfaces is governed by the precession ellipticity, which acts as a magnonic temporal impedance $Z_i \equiv \varepsilon^z_i$ in direct analogy to electromagnetic wave impedance at dielectric boundaries. Any temporal impedance mismatch universally expands the precession orbit, and adiabatic field ramps suppress temporal reflections exponentially following Landau--Zener dynamics, enabling wavepacket manipulation without back-scattering.

Second, near the field-driven stripe-domain transition, a damping-induced instability emerges in the slow-instability window $H_{\mathrm{EP}} < H < H_c$, opened by DMI-induced nonreciprocity. In this regime the spin-wave growth rate scales linearly with the Gilbert damping parameter---a counterintuitive behavior where increased dissipation enhances rather than suppresses gain, naturally interpreted as amplification driven by antimagnon modes generated through time-refraction at the first temporal interface.

The temporal-slab protocol synthesizes both foundations: adiabatic interfaces provide reflection-free entry and exit, while the slow-instability regime accumulates exponential gain during the low-field plateau. This combination achieves frequency-preserving amplification up to two orders of magnitude without lithographic patterning---substantially exceeding existing parametric and spin-torque  (see Table~S2 in Supplemental Information). 
The energy for amplification is supplied by finite work performed during the field transition rather than by continuous resonant pumping: the field ramp prepares a metastable uniform state below $H_c$, whose excess energy is subsequently released into growing antimagnonic spin-wave excitations. This distinguishes the temporal-slab mechanism from conventional parametric drives and links it to the recently introduced antimagnonic framework.
The analytical expressions derived here are applicable to any PMA system with field-driven stripe-domain transitions and sufficient DMI. Combined with the demonstrated temporal-slab protocol, these results show that temporal field modulation can serve as a practical tool for spin-wave amplification, frequency conversion, and wavevector-selective filtering in thin-film magnetic systems.

\section{Methods}
    \subsection{Mathematical methods}\label{app:model}

        For completeness, extended derivations and intermediate steps are provided in the Supplemental Information~\cite{SM}.

        We consider an ultrathin ferromagnetic film (thickness $d$) with equilibrium magnetization $\mathbf{m}_0 = (0, M_s, 0)$ under an external magnetic field $\mathbf{H}_0 = H_{1,2} \hat{\mathbf{y}}$. Small-amplitude spin-wave dynamics are described by deviations $\mathbf{m} = (m_x, M_s, m_z)$ with $m_x, m_z \ll M_s$.

            The linearized Landau--Lifshitz equation for small deviations from equilibrium ($m_x, m_z \ll M_s$)~\cite{GurevichMelkov1996} yields (neglecting damping):
            \begin{align}
                \partial_t m_x &= \omega_x m_z + i\, s\omega_D m_x, \\
                \partial_t m_z &= -\omega_z m_x + i\, s\omega_D m_z,
            \end{align}
            where $s = \mathrm{sgn}(v_\mathrm{ph})$ accounts for DMI-induced nonreciprocity. The characteristic frequencies are
            \begin{align}
                \omega_x &= A\bigl[h + l_\mathrm{ex}^2 k^2 - Q + 1 - \xi(|k|d)\bigr],
                \label{eq:omega_x} \\
                \omega_z &= A\bigl[h + l_\mathrm{ex}^2 k^2 + \xi(|k|d)\bigr],
                \label{eq:omega_z} \\
                \omega_D &= A l_D |k|.
                \label{eq:omega_D}
            \end{align}
            
            where $A = \gamma \mu_0 M_s$, $h = H/M_s$, $Q = K_\mathrm{PMA}/(\tfrac{1}{2}\mu_0 M_s^2)$ parameterizes perpendicular magnetic anisotropy, $l_\mathrm{ex} = \sqrt{2A_\mathrm{ex}/(\mu_0 M_s^2)}$ and $l_D = 2D/(\mu_0 M_s^2)$ are exchange and DMI lengths, and $\xi(|k|d) = 1 - (1 - e^{-|k|d})/(|k|d)$ accounts for dipolar interactions.
            

            Substituting the plane-wave ansatz in time $m_{x,z} \propto \exp(-i\Omega t)$ leads to the dispersion relation
            \begin{equation}
                \Omega_\pm(k) = -s\omega_D(k) \pm \sqrt{\omega_x(k)\,\omega_z(k)},
                \label{eq:dispersion_undamped}
            \end{equation}
            with two branches corresponding to counterpropagating waves.
            The physical branch (positive frequency for $k > 0$) is $\Omega_+ = \sqrt{\omega_x \omega_z} - s\omega_D$.
            The associated eigenvectors of the linearized dynamics are
            \begin{equation}
                \begin{pmatrix} m_x \\ m_z \end{pmatrix}_\pm \propto \begin{pmatrix} 1 \\ \mp\,i\,\varepsilon^z \end{pmatrix}, \qquad
                \varepsilon^z \equiv \frac{|m_z|}{|m_x|} = \sqrt{\frac{\omega_z}{\omega_x}},
                \label{eq:eigenvectors_undamped}
            \end{equation}
            so that $m_z = -i\varepsilon^z m_x$ on the $\Omega_+$ branch and $m_z = +i\varepsilon^z m_x$ on the $\Omega_-$ branch; the modulus $\varepsilon^z$ defines the precession ellipticity, and the opposite phase between the two branches encodes the chirality of precession that distinguishes counterpropagating waves.
            A full derivation is given in Sec.~S1.2 of the Supplemental Information.

        A sudden change in the bias field $h_1 \to h_2$ at time $t_0$ creates a temporal interface. Physical continuity of magnetization requires:
        \begin{equation}
            m_x(t_0^-) = m_x(t_0^+), \qquad m_z(t_0^-) = m_z(t_0^+).
            \label{eq:continuity}
        \end{equation}
        Since the spatial profile cannot change instantaneously, the wavevector $k$ remains constant while the frequency adjusts to the new dispersion.
        
        For an incident rightward-propagating wave with amplitude $A_1$, 
        two waves emerge in medium 2: transmitted/time-refracted ($\Omega_2^{(t)} > 0$, amplitude $A_t$) and time-reflected
        ($\Omega_2^{(r)} < 0$, amplitude $A_r$). Crucially, the two dispersion branches $\Omega_{\pm} = -s\omega_D \pm \sqrt{\omega_x\omega_z}$ have different phase relationships between magnetization components. From the eigenvalue problem, the ratio $m_z/m_x \propto (\Omega + s\omega_D)$, which gives:
        \begin{equation}
            m_z = -i\varepsilon^z m_x \quad \text{for } \Omega_+, \qquad 
            m_z = +i\varepsilon^z m_x \quad \text{for } \Omega_-.
        \end{equation}
        The refracted wave lies on the $\Omega_+$ branch, while the reflected wave lies on the $\Omega_-$ branch.
        
        Applying continuity conditions~\eqref{eq:continuity}:
        \begin{align}
            m_x: \quad A_1 &= A_t + A_r, \\
            m_z: \quad -i\varepsilon^z_1 A_1 &= -i\varepsilon^z_2 A_t + i\varepsilon^z_2 A_r,
        \end{align}
        where the opposite signs in the $m_z$ equation reflect the different branches. Simplifying yields $\varepsilon^z_1 A_1 = \varepsilon^z_2 (A_t - A_r)$. Solving for $T_x = A_t/A_1$ and $R_x = A_r/A_1$:
        \begin{equation}
            T_x = \frac{1}{2}\left(1 + \frac{\varepsilon^z_1}{\varepsilon^z_2}\right), \quad R_x = \frac{1}{2}\left(1 - \frac{\varepsilon^z_1}{\varepsilon^z_2}\right).
        \end{equation}
        The $m_z$ coefficients follow from $T_z = m_z^{(t)}/m_z^{(1)} = (\varepsilon^z_2/\varepsilon^z_1) T_x$ and $R_z = m_z^{(r)}/m_z^{(1)} = -(\varepsilon^z_2/\varepsilon^z_1) R_x$, where the minus sign arises because the reflected wave has opposite $m_z/m_x$ phase. This yields Eqs.~\eqref{eq:T_R_z_eps}.

    By analogy with photonic and electronic impedances, we define the magnonic temporal impedance $Z_i \equiv \varepsilon^z_i$. In this notation, the transmission and reflection coefficients become:
    \begin{equation}
    T_x = \frac{1}{2}\left(1 + \frac{Z_1}{Z_2}\right), \quad R_x = \frac{1}{2}\left(1 - \frac{Z_1}{Z_2}\right),
    \label{eq:T_R_impedance_x}
    \end{equation}
    \begin{equation}
    T_z = \frac{1}{2}\left(1 + \frac{Z_2}{Z_1}\right), \quad R_z = \frac{1}{2}\left(1 - \frac{Z_2}{Z_1}\right).
    \label{eq:T_R_impedance_z}
    \end{equation}
    The precession ellipse area transmission coefficient is $T_S = T_x \cdot T_z$, which always exceeds unity when $Z_1 \neq Z_2$, indicating a universal orbit expansion at temporal interfaces.

    Detailed derivations and additional validation are provided in the Supplemental Information~\cite{SM}.

        Incorporating Gilbert damping into the linearized Landau-Lifshitz-Gilbert equations yields:
        \begin{equation}
            \partial_t m_x = \omega_x m_z + i\,  s \omega_D m_x + \alpha \partial_t m_z,
            \label{eq:LLG_damped_mx}
        \end{equation}
        \begin{equation}
            \partial_t m_z = -\omega_z m_x + i\,  s \omega_D m_z - \alpha \partial_t m_x,
            \label{eq:LLG_damped_mz}
        \end{equation}
        where $\alpha$ is the Gilbert damping parameter, $s = \mathrm{sgn}(v_\mathrm{ph})$, and $\omega_x$, $\omega_z$ are the characteristic frequencies defined as $\omega_{x,z} = \gamma \mu_0 H_{\mathrm{eff};x,z}$.
        Substituting the plane-wave ansatz in time, $m_{x,z} \propto \exp(-i\Omega t)$, the condition for nontrivial solutions leads to a quadratic equation with \textit{complex coefficients}:
        \begin{equation}
            (1 + \alpha^2)\Omega^2 + [2s\omega_D + i\alpha(\omega_x + \omega_z)]\Omega + (\omega_D^2 - \omega_x\omega_z) = 0.
            \label{eq:quadratic_dispersion}
        \end{equation}
        
        The solution $\Omega_{\pm} = \mathrm{Re}(\Omega_{\pm}) + i\,\mathrm{Im}(\Omega_{\pm})$ is given by:
        \begin{equation}
            \mathrm{Re}(\Omega_{\pm}) = \frac{-2s\omega_D \pm u}{2(1+\alpha^2)}, \qquad 
            \mathrm{Im}(\Omega_{\pm}) = \frac{-\alpha(\omega_x+\omega_z) \pm v}{2(1+\alpha^2)},
            \label{eq:ReIm_Omega}
        \end{equation}
        where $u$ and $v$ are determined by the complex discriminant $\Delta = X + iY$:
        \begin{equation}
            u = \sqrt{\frac{|\Delta| + X}{2}}, \qquad v = \mathrm{sgn}(Y)\sqrt{\frac{|\Delta| - X}{2}},
        \end{equation}
        with $|\Delta| = \sqrt{X^2 + Y^2}$ and
        \begin{align}
            X &= 4(1+\alpha^2)\omega_x\omega_z - \alpha^2[(\omega_x+\omega_z)^2 + 4\omega_D^2], \\
            Y &= 4s\alpha\omega_D(\omega_x+\omega_z).
        \end{align}
        
        The critical frequency separating dynamical regimes is:
        \begin{equation}
            \omega_x^* = \frac{\omega_D^2}{\omega_z},
            \label{eq:omega_x_critical}
        \end{equation}
        corresponding to the condition $\mathrm{Re}(\Omega) = 0$ at the onset of the magnetic phase transition. The exceptional point (see proof in SM) occurs at $\omega_x = 0$, where the two branches coalesce for $\alpha = 0$ with characteristic splitting $|\Omega_+ - \Omega_-| \propto \sqrt{\alpha}$ for finite damping. The physical interpretation of the resulting dynamical regimes is discussed in the Results section and summarized in Table~\ref{tab:regimes_results}.
        
    For small Gilbert damping ($\alpha \ll 1$, typically $\alpha \lesssim 0.01$), the imaginary part of frequency exhibits distinct scaling in each regime. For larger damping ($\alpha \sim 0.1$), corrections of order 10--20\% may apply, but the qualitative conclusions remain unchanged.
    
    In the \textbf{strong instability regime} ($\omega_x < 0$):
    \begin{equation}
        \mathrm{Im}(\Omega_+) \approx \sqrt{|\omega_x|\omega_z} + \frac{\alpha(|\omega_x| - \omega_z)}{2}.
        \label{eq:Im_strong_instability}
    \end{equation}
    The growth rate is dominated by the magnetic instability term $\sqrt{|\omega_x|\omega_z}$, which is independent of $\alpha$. The correction term is linear in $\alpha$ and typically small (a few percent for $\alpha \sim 0.01$).
    
    For $\omega_x > 0$, encompassing both the \textbf{slow instability} and \textbf{damping} regimes, a unified formula applies:
    \begin{equation}
    \begin{aligned}
        \mathrm{Im}(\Omega_+) \approx& \frac{\alpha(\omega_x + \omega_z)}{2}\left[\frac{\omega_D}{\sqrt{\omega_x\omega_z}} - 1\right] = \frac{\alpha(\omega_x + \omega_z)}{2}\left[\sqrt{\frac{\omega_x^*}{\omega_x}} - 1\right].
    \end{aligned}
        \label{eq:Im_unified}
    \end{equation}
    This expression is positive (amplification) when $\omega_x < \omega_x^*$ and negative (damping) when $\omega_x > \omega_x^*$, with the growth/decay rate scaling \textit{linearly} with $\alpha$ in both cases. The accuracy of these approximations is validated in Supplemental Information, Fig.~S1~\cite{SM}.
    
    Specifically, in the \textbf{slow instability regime} ($0 < \omega_x < \omega_x^*$):
    \begin{equation}
        \mathrm{Im}(\Omega_+) > 0, \quad \text{with rate} \propto \alpha.
        \label{eq:Im_slow_instability}
    \end{equation}
    Amplification vanishes in the conservative limit $\alpha \to 0$.
    
    In the \textbf{damping regime} ($\omega_x > \omega_x^*$):
    \begin{equation}
        \mathrm{Im}(\Omega_+) < 0, \quad \text{with rate} \propto \alpha.
        \label{eq:Im_damping}
    \end{equation}

    The regime boundaries in terms of normalized bias field $h = H_0/M_s$ are:
    (Figs.~\ref{fig:fig_5}g and~\ref{fig:k_selectivity}d)
    \begin{equation}
        h_{\mathrm{EP}}(k) = Q - 1 + \xi(k) - l_{\mathrm{ex}}^2 k^2,
        \label{eq:h_EP}
    \end{equation}
    \begin{equation}
        h_c(k) = h_{\mathrm{EP}}(k) + \frac{\omega_D^2(k)}{\gamma\mu_0 M_s \, \omega_z(k)},
        \label{eq:h_c}
    \end{equation}
    where $h_{\mathrm{EP}}$ corresponds to $\omega_x = 0$ (exceptional point) and $h_c$ to $\omega_x = \omega_x^*$ (critical field, phase transition onset). The amplification window width:
    \begin{equation}
        \Delta h = h_c - h_{\mathrm{EP}} =  \frac{\omega_D^2}{\gamma\mu_0 M_s \, \omega_z} = \frac{4 D^2 k^2}{\mu_0^2 M_s^4 \bigl[h + l_{\mathrm{ex}}^2 k^2 + \xi(k)\bigr]},
        \label{eq:Delta_h}
    \end{equation}
    scales as $\Delta h \propto D^2/M_s^4$ or as $\Delta H \propto D^2/M_s^3$ in dimensional units ($\Delta H = M_s \Delta h$) and is independent of $\alpha$, while the growth rate within this window scales linearly with $\alpha$.    

     The soft-mode wavevector $k_{\mathrm{soft}}$ corresponds to the maximum of $h_c(k)$, i.e., the wavevector at which the dispersion first reaches zero frequency upon lowering the field. In the limit of weak DMI and ultrathin films ($kd \ll 1$), the condition $\partial h_c / \partial k = 0$ reduces to $\partial \xi / \partial k = 2 l_{\mathrm{ex}}^2 k$, yielding:
    \begin{equation}
        k_{\mathrm{soft}} \approx \frac{d}{4 l_{\mathrm{ex}}^2}.
        \label{eq:k_soft}
    \end{equation}
    For our CoFeB parameters ($d = 2$~nm, $l_{\mathrm{ex}} = 3.2$~nm), this gives $k_{\mathrm{soft}} \approx 49$~rad/\textmu m, in agreement with micromagnetic simulations (Fig.~\ref{fig:fig_2}a).
    Throughout the main text, $H_{\mathrm{EP}}$ denotes the maximum of $H_{\mathrm{EP}}(k)$, reached at $k=k_{\mathrm{EP}}$, whereas $H_c$ denotes the maximum of $H_c(k)$, reached at $k=k_{\mathrm{soft}}$. The amplification window is therefore $\Delta H = H_c - H_{\mathrm{EP}}$.

    \begin{equation}
        H_c \equiv H_c(k_{\mathrm{soft}}), \quad
        H_{\mathrm{EP}} \equiv H_{\mathrm{EP}}(k_{\mathrm{EP}}), \quad
        \Delta H \equiv H_c(k_{\mathrm{soft}})-H_{\mathrm{EP}}(k_{\mathrm{EP}}).
    \end{equation}

    \subsection{Micromagnetic simulations}\label{app:simulations}

        Numerical simulations were performed using Mumax3 \cite{vansteenkiste2014design} to solve the full Landau-Lifshitz-Gilbert equation. We simulated a CoFeB film with thickness $d = 2$ nm and the following material parameters: saturation magnetization $M_s = 1420$ kA/m, exchange stiffness $A_\mathrm{ex} = 13$ pJ/m, DMI strength $D = 0.5$ mJ/m$^2$, damping parameter $\alpha = 0.002$ (unless stated otherwise), and reduced anisotropy constant $Q = 1.1$, where $Q = \tfrac{K_\mathrm{PMA}}{\frac{1}{2}\mu_0 M_s^2}$ with $K_\mathrm{PMA}$ being the uniaxial anisotropy constant. The system was discretized with unit cells of size $3 \times 3 \times 2$ nm$^3$ along the $x$-, $y$-, and $z$-directions respectively. The simulated geometry comprised a length of 90 \textmu m and width of $30$~nm with periodic boundary conditions applied along the $x$- and $y$-directions.
        To validate that the quasi-one-dimensional geometry does not introduce artifacts, we performed additional simulations with 1000 cells along the $y$-direction (instead of a single cell with periodic conditions) and obtained identical results, confirming that the relevant physics is effectively one-dimensional for the propagation geometry considered here.
        
        Each simulation began with a uniform magnetic configuration along the $y$-direction, which was subsequently relaxed to equilibrium configuration under a static magnetic field of $300$~mT applied along the $y$-direction. The dynamic simulations were performed with a locally applied microwave magnetic field of spatial and temporal profiles designed first to obtain the dispersion relation and later to excite propagating wavepackets.
        
        For simulations with a single temporal interface, we employed the following time dependence of the $y$-component of the external magnetic field Eq.~\eqref{eq:singleSmoothInterface}.
        The step-function limit was obtained by setting the limit $\tau \to 0$.
        For the simulations with two temporal interfaces forming a temporal slab, we used Eq.~\eqref{eq:temporal_slab}.
        In all simulations, we set the location of the center of first temporal interface at $t_i = 10$~ns.
        
        To compute the dispersion relation, we used the spatial and temporal dependences of the out-of-plane applied microwave field
        \begin{equation}
        h_{\mathrm{exc},z}(t,x) = h_0 \mathrm{sinc}(k_\mathrm{c}x) \mathrm{sinc}(2 \pi f_\mathrm{c}(t - t_0)),
        \end{equation}
        where $t_0=10/f_c$, the cutoff parameters $k_\mathrm{c} = 100$~rad/\textmu m and $f_\mathrm{c} = 8$~GHz define the range of excited wavenumbers and frequencies. The simulation results were recorded with the time step of $(2.2 f_\mathrm{c})^{-1}$. The dispersion relation was obtained by computing the two-dimensional FFT in time and space of the out-of-plane magnetization component $m_z$: $|\tilde{M}_z|(f, k_x) = \mathrm{FFT}_{t,x}(m_z)$. For each dispersion plot, the color scale was normalized to maximize the readability. To compare dispersions for different magnetic field values (as shown in Figs.~\ref{fig:fig_1}(c,d) and~\ref{fig:fig_2}a), each FFT result was mapped to one or more RGB color channels, enabling composite visualization of multiple field configurations in a single image.
        
        For simulations of spin-wave wavepackets propagating in a selected direction, we employed the following spatial and temporal dependence of the out-of-plane applied microwave field\cite{whitehead2019graded}:
        \begin{equation}
        \begin{aligned}
        h_{\mathrm{exc}, z}(t,x) =& h_{\mathrm{exc},0}G(t)G(x)[\sin(k x)\sin(2 \pi f_0 t) + \cos(k x)\cos(2 \pi f_0 t)],
        \end{aligned}
        \end{equation}
        where $h_{\mathrm{exc},0}$ is the peak amplitude of the excitation field (typically $\mu_0 h_{\mathrm{exc},0} = 20$ \textmu T), $G(t) = \exp[-2.77(t-t_0)^2/(2\sigma_t^2)]$ with $\sigma_t = 1/f_0$ provides the temporal envelope centered at $t_0=4T$ ($T$ being one period of microwave field oscillation), and $G(x) = \exp[-x^2/(2\sigma_x^2)]$ with $\sigma_x = 40\pi/k$ defines the spatial profile. The envelope parameters correspond to FWHM in space and time domains of approximately $w_t=2.355/f_0$ and $w_x=295.6/k$ (or $w_x=47.1\,\lambda$ wavelengths), respectively. The wavenumber $k$ and frequency $f_0$ were determined from the dispersion relations obtained in the previous simulations. The simulation results of wavepacket scattering were recorded with the time step of $(4 f_0)^{-1}$.
        
        To analyze the spatiotemporal propagation of spin-wave wavepackets, we processed the micromagnetic simulation results of the in-plane and out-of-plane magnetization component $m_{x,z}(t,x)$ using the Hilbert transform. At each time instant, the Hilbert transform was applied to extract the wavepacket envelope. To maximize the accuracy of the envelope representation, we applied Fourier filtering to remove high-frequency spatial components of wavelength smaller than $10$~nm, and adjusted the envelope maximum to align with the actual wavepacket maximum. This refinement was justified because the wavepackets were significantly broader than the wavelength. Using this procedure, we tracked both the trajectory (center of mass) and the FWHM of the wavepacket envelope with high precision. Representative examples of wavepacket propagation and envelope evolution are shown in Figs.~\ref{fig:fig_1}--\ref{fig:fig_6}.

        To extract transmission and reflection coefficients from the micromagnetic simulations, we separated incident, refracted, and reflected wavepackets using two-dimensional Fourier analysis of the magnetization dynamics $m(x,t)$.
        
        The power spectrum in frequency-wavevector space reveals two branches corresponding to rightward ($k > 0$ at $f>0$ and $k<0$ at $f<0$) and leftward ($k < 0$ at $f>0$ and $k>0$ at $f<0$) propagating modes. Due to nonreciprocal dispersion, counterpropagating modes at $\pm|k|$ have different frequencies. We identified the dominant wavevector $|k|$ from the integrated power spectrum, then extracted frequency profiles at $k = \pm|k|$ using a narrow integration window. Peak detection with Gaussian smoothing identified the characteristic frequencies for each propagation direction.
        
        Wavepackets were separated by applying spectral masks to the Fourier-transformed data. Each mask consisted of two Gaussian envelopes centered at $(f_i, +|k|)$ and $(-f_i, -|k|)$ with widths $\sigma_f = 10$ and $\sigma_k = 5$ frequency and wavevector points, respectively. Inverse FFT of the masked spectra yielded complex wavepackets amplitude $\psi(x,t)$.
        
        The envelope amplitudes were obtained as spatial maxima of $|\psi(x,t)|$ at each simulation time step and fitted to exponential functions $A + B\exp(-t/\eta)$ in the quasi-steady-state regime ($t \gg t_i$). This approach avoids artifacts from the FFT boundary effects due to assumed temporal periodicity. Transmission and reflection coefficients were calculated by extrapolating these fits to the interface time $t_i$ and computing amplitude ratios immediately after and before the temporal interface.

        For gradual temporal interfaces (Fig.~\ref{fig:fig_4}j), the dispersion-based FFT filtering approach was unsuccessful, as gradual field transitions do not produce sharp, well-defined modes in the dispersion relation. Instead, we employed a direct fitting approach: the time-dependent amplitude envelope was fitted to a damped oscillation model superimposed on a polynomial background:
        \begin{equation}
        A(t) = A_{\mathrm{osc}} e^{-t/\tau} \cos(\omega t + \phi) + P(t),
        \end{equation}
        where $A_{\mathrm{osc}}$ is the oscillation amplitude, $\tau$ is the decay time, $\omega$ and $\phi$ are the frequency and phase, and $P(t)$ is a polynomial background. From the fitted parameters $A_{\mathrm{osc}}$ and the background value $B = P(0)$, the transmission and reflection coefficients were calculated using:
        \begin{equation}
        \begin{aligned}
        |t| &= \frac{\sqrt{B + A_{\mathrm{osc}}} + \sqrt{B - A_{\mathrm{osc}}}}{2},\\ 
        |r| &= \frac{\sqrt{B + A_{\mathrm{osc}}} - \sqrt{B - A_{\mathrm{osc}}}}{2}.
        \end{aligned}
        \end{equation}

    \subsection{Use of AI tools}

        The authors used Claude Opus 4.5 (Anthropic) to improve English grammar and clarity, and as an interactive tool during the development of analytical derivations---including verification of mathematical steps and evaluation of approximations. During these discussions, AI suggested interpreting the degeneracy at $\omega_x = 0$ as an exceptional point with characteristic $\sqrt{\alpha}$ splitting, and provided a draft proof of this identification. The authors independently verified this interpretation through analytical derivation. All simulations, calculations, and scientific conclusions were performed by the authors.

\section*{Data availability}
All data needed to evaluate the conclusions are present in the paper and/or the Supplemental Informations. MuMax3 input files are available at Zenodo (DOI: 10.5281/zenodo.17864084). Custom analysis scripts are available from the corresponding authors.

\bibliographystyle{unsrt}
\bibliography{literature_v2}

\section*{Acknowledgments}
The research leading to these results has received funding from the Polish National Science Centre projects No. 2019/35/D/ST3/03729 and 2022/45/N/ST3/01844.
This work was carried out using the infrastructure of the Poznań Supercomputing and Networking Center (PCSS), within Scientific Computing Space no. pl0804.

\paragraph*{Author contributions:} P.G. conceived the project and developed the analytical framework.  K.S. developed the simulation codes, performed the micromagnetic simulations, and designed the wave-packet separation algorithm.  P.G. conducted the theoretical analysis and prepared the Supplementary Information including the simulations presented therein.  P.G. wrote the manuscript, prepared the responses to the reviewers, and revised the manuscript accordingly, with the input of K.S.  Both authors discussed the results and revised the manuscript.  P.G. supervised the project.  P.G. and K.S. independently acquired funding.

\paragraph*{Competing interests:} The authors declare no competing financial or non-financial interests.


\newpage

\end{document}